\documentclass[10pt, amsmath, amssymb, preprintnumbers, showpacs,showkeys,aps,prb,superscriptaddress,twocolumn,longbibliography]{revtex4-1}

\pdfoutput=1 
\usepackage{amssymb}
\usepackage{dsfont}
\usepackage[utf8]{inputenc}
\usepackage{graphicx}
\usepackage{amsmath}
\usepackage{csquotes}
\usepackage{xcolor}
\usepackage[normalem]{ulem} 
\usepackage{physics}
\usepackage[pdftex,colorlinks=true,citecolor=cyan,linkcolor=magenta]{hyperref}
\usepackage[capitalize]{cleveref}
\usepackage{newtxtext}
\usepackage{siunitx}
\usepackage{orcidlink} 
\usepackage{placeins}
\usepackage{titlesec}

\usepackage{natbib}

\usepackage{bibunits}
\defaultbibliography{bibliography}
\defaultbibliographystyle{naturemag}

\let\oldaddcontentsline\addcontentsline     
\renewcommand{\addcontentsline}[3]{}  

\def\sectitle#1{\smallskip\medskip {\noindent \fontsize{11pt}{12pt}\selectfont {\textbf{#1}}\noindent }}

\def\subsectitle#1{\vspace{1.5mm}

{\noindent{\textbf{#1}}}}

\graphicspath{{SI/}} 

\newcounter{myfigure}
\makeatletter
\@addtoreset{figure}{myfigure}
\makeatother

\renewcommand\thefigure{{\textbf{{\arabic{figure}}}}}
\newcommand{\titledcaption}[2]{\caption{\textsf{{\textbf{#1}#2}}}}

\newcommand{\phasefront}{Supplementary Movie~1}

\newcommand{\e}{\mathrm{e}}         
\renewcommand{\i}{\mathrm{i}}       
\def\abs#1{\left|{#1}\right|}   

\newcommand{\adjo}[1]{{#1}^\dagger}
\newcommand{\cconj}[1]{{#1}^*}
\newcommand{\id}{\mathds{1}}
\renewcommand{\vec}[1]{\mathbf{#1}}

\DeclareMathOperator\arctanh{arctanh}

\newcommand{\nunit}[1]{\,\mathrm{#1}}
\DeclareSIUnit\bit{bit}

\newcommand{\eqcontrib}{These two authors contributed equally to this work.}
\newcommand{\comms}{Correspondence to \href{mailto:titus.neupert@uzh.ch}{titus.neupert@uzh.ch}, \href{rthomale@physik.uni-wuerzburg.de}{rthomale@physik.uni-wuerzburg.de}. and \href{mailto:tomas.bzdusek@psi.ch}{tomas.bzdusek@psi.ch}.}

\begin{document}
\begin{bibunit}

\title{Simulating hyperbolic space on a circuit board}

\author{Patrick M. Lenggenhager\,\orcidlink{0000-0001-6746-1387}}\thanks{\eqcontrib{}}
\affiliation{Condensed Matter Theory Group, Paul Scherrer Institute, 5232 Villigen PSI, Switzerland}
\affiliation{Department of Physics, University of Zurich, Winterthurerstrasse 190, 8057 Zurich, Switzerland}
\affiliation{Institute for Theoretical Physics, ETH Zurich, 8093 Zurich, Switzerland}

\author{Alexander Stegmaier\,\orcidlink{0000-0002-8864-5182}}\thanks{\eqcontrib{}}
\affiliation{Institut für Theoretische Physik und Astrophysik, Universität Würzburg, 97074 Würzburg, Germany}

\author{Lavi K. Upreti\,\orcidlink{0000-0002-1722-484X}}
\affiliation{Institut für Theoretische Physik und Astrophysik, Universität Würzburg, 97074 Würzburg, Germany}

\author{Tobias Hofmann\,\orcidlink{0000-0002-1888-9464}}
\affiliation{Institut für Theoretische Physik und Astrophysik, Universität Würzburg, 97074 Würzburg, Germany}

\author{Tobias Helbig\,\orcidlink{0000-0003-1894-0183}}
\affiliation{Institut für Theoretische Physik und Astrophysik, Universität Würzburg, 97074 Würzburg, Germany}

\author{Achim Vollhardt\,\orcidlink{0000-0002-4424-1127}}
\affiliation{Department of Physics, University of Zurich, Winterthurerstrasse 190, 8057 Zurich, Switzerland}

\author{Martin Greiter}
\affiliation{Institut für Theoretische Physik und Astrophysik, Universität Würzburg, 97074 Würzburg, Germany}

\author{Ching Hua Lee}
\affiliation{Department of Physics, National University of Singapore, Singapore 117551, Republic of Singapore}

\author{Stefan Imhof}
\affiliation{Physikalisches Institut, Universität Würzburg, 97074 Würzburg, Germany}

\author{Hauke Brand\,\orcidlink{0000-0003-4209-2663}}
\affiliation{Physikalisches Institut, Universität Würzburg, 97074 Würzburg, Germany}

\author{Tobias Kie\ss ling}
\affiliation{Physikalisches Institut, Universität Würzburg, 97074 Würzburg, Germany}

\author{Igor Boettcher\,\orcidlink{0000-0002-1634-4022}}
\affiliation{Department of Physics, University of Alberta, Edmonton, Alberta T6G 2E1, Canada}
\affiliation{Theoretical Physics Institute, University of Alberta, Edmonton, Alberta T6G 2E1, Canada}

\author{Titus Neupert\,\orcidlink{0000-0003-0604-041X}}\thanks{\comms{}}
\affiliation{Department of Physics, University of Zurich, Winterthurerstrasse 190, 8057 Zurich, Switzerland}

\author{Ronny Thomale\,\orcidlink{0000-0002-3979-8836}}\thanks{\comms{}}
\affiliation{Institut für Theoretische Physik und Astrophysik, Universität Würzburg, 97074 Würzburg, Germany}

\author{Tom\'{a}\v{s} Bzdu\v{s}ek\,\orcidlink{0000-0001-6904-5264}}\thanks{\comms{}}
\affiliation{Condensed Matter Theory Group, Paul Scherrer Institute, 5232 Villigen PSI, Switzerland}
\affiliation{Department of Physics, University of Zurich, Winterthurerstrasse 190, 8057 Zurich, Switzerland}

\date{\today}

\begin{abstract}
\pdfbookmark[0]{Abstract}{abstract}\sectitle{Abstract}

\noindent The Laplace operator encodes the behaviour of physical systems at vastly different scales, describing heat flow, fluids, as well as electric, gravitational, and quantum fields. A key input for the Laplace equation is the curvature of space. Here we discuss and experimentally demonstrate that the spectral ordering of Laplacian eigenstates for hyperbolic (negatively curved) and flat two-dimensional spaces has a~universally different structure. We use a~lattice regularization of hyperbolic space in an electric-circuit network to measure the eigenstates of a~`hyperbolic drum', and in a time-resolved experiment we verify signal propagation along the curved geodesics. Our experiments showcase both a versatile platform to emulate hyperbolic lattices in tabletop experiments, and a set of methods to verify the effective hyperbolic metric in this and other platforms. The presented techniques can be utilized to explore novel aspects of both classical and quantum dynamics in negatively curved spaces, and to realise the emerging models of topological hyperbolic matter.
\end{abstract}

\maketitle

\pdfbookmark[0]{Introduction}{introduction}\sectitle{Introduction}

\noindent Curved spaces, traditionally studied in high-energy physics and cosmology, have recently been elevated to paramount importance in condensed matter physics for two reasons. 
First, the discovery of holographic principles~\cite{Maldacena:1999,Witten:1998} revealed a fundamental hidden structure underlying certain interacting quantum many-body systems that allows to compute their properties from a theory in hyperbolic space of negative curvature. 
Remarkably, these insights have been applied successfully to analyze strongly correlated electronic systems with tools from holography and to gain insight into the nature of quantum entanglement in condensed matter systems~\cite{Hartnoll:2018,Ryu:2006,Vidal:2007,Son:2008,Vidal:2008,Matsueda:2013,Swingle:2012,Haegeman:2013,Boyle:2020}. 
Second, major advancements in the mathematical characterization of classical and quantum states in negatively curved spaces~\cite{Maciejko:2021,Maciejko:2020,Ikeda:2021,Boettcher:2021} sparked a resurgence of interest of the condensed matter and metamaterials communities in hyperbolic lattices~\cite{Kollar:2019,Boettcher:2020,Asaduzzaman:2020}, ushering the research of hyperbolic topological matter~\cite{Yu:2020,Urwyler:2021,Bienias:2021}.
These rapid developments call for new experimental platforms to implement tabletop simulations of hyperbolic toy-models.

However, systems that furnish negatively curved space~\cite{Coxeter:1957,Coxeter:1979} are hard to realise experimentally.
The mathematical reason for this is encompassed in Hilbert's theorem:
even the lowest dimensional model of a~hyperbolic space, the hyperbolic plane, cannot be embedded in three-dimensional Euclidean (flat) laboratory space. 
We cannot build a~hyperbolic drum.
This is in sharp contrast to the case of positive curvature:
a sphere can be embedded in three-dimensional space, and we can study the standing waves (hereafter called eigenmodes) of a~spherical membrane, which directly relate to quantum numbers of atomic orbitals.
Despite such obstacles, hyperbolic space can be emulated experimentally.
For instance, it has been suggested~\cite{Leonhardt:2006} that a~non-trivial metric can be implemented in metamaterials via spatial variations of the electromagnetic permittivity of continuous media.
However, it is very challenging to induce these variations in a~controlled manner, which limits the applicability of such approaches.

Electric circuits~\cite{Cserti:2000,Cserti:2011,Ningyuan:2015,Albert:2015,Lee:2018,Imhof:2018,Helbig:2019,Hofmann:2020} and similar systems, e.g., coplanar waveguide resonators~\cite{Kollar:2019}, overcome these experimental limitations by relying on a~discretization of space.
In electric circuit networks, the physical distances between the nodes are fundamentally decoupled from the metric that enters the long-wavelength description of its degrees of freedom, namely the voltages and currents that pass through the circuit nodes.
The latter depend merely on the circuit elements that connect the nodes.
Compared to other experimental platforms, electric circuits significantly excel in their flexibility of design, ease of fabrication, and high accessibility to measurements.

\begin{figure*}[t]
    \centering
    \includegraphics{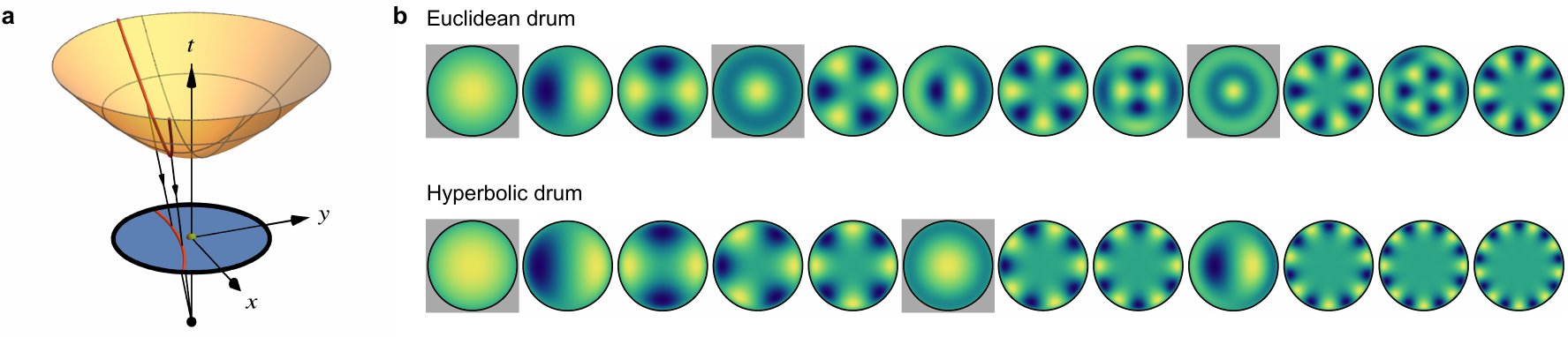}
    \titledcaption{Continuum spectra.}{
    \textbf{a} The hyperboloid (orange) defined by $t^2 - x^2 - y^2 = +1$ in (2+1)-dimensional $(x,y,t)$ Minkowski space is mapped (black rays) by the stereographic projection through the point $(0,0,-1)$ (black dot) to the unit disk (blue) at $t=0$. The geodesics (red) are given by intersections of the hyperboloid with planes passing through the origin $(0,0,0)$ (green dot), and are mapped by the projection to circular arcs perpendicular to the boundary of the Poincaré disk.
    \textbf{b} Comparison of the first few eigenmodes of the Euclidean and hyperbolic drum of radius $r_0=0.94$ according to increasing eigenvalues $\lambda_\mathrm{g}^{n \ell}$. Their spatial profile $u_\mathrm{g}^{n \ell}$ is shown with yellow (green, blue) denoting maxima (zeros, minima). The number of radial zeros inside the disk, $n$, and the angular momentum (number of angular zeroes), $\ell$,  can easily be inferred from the plots. Modes with $\ell=0$ are indicated with a~grey background.
    }
    \label{fig:continuum}
\end{figure*}

In this work we present a strategy for verifying that electric circuits can emulate the physics of negatively curved spaces and we demonstrate that electric circuits can do so efficiently.
For concreteness, we consider the most fundamental differential operator on curved spaces, the Laplace-Beltrami operator, which generalizes the notion of the Laplace operator on flat space.
The first key result of our work is the experimental observation of negative curvature in the spectral ordering of the eigenmodes of the Laplace-Beltrami operator in hyperbolic space.
To paraphrase the words of Ref.~\onlinecite{Kac:1966}, our measurements confirm that a~hyperbolic drum has a~sound distinct from a~Euclidean drum.
Second, since electric circuits allow for time-resolved measurements, we can study not only static, but also dynamic properties.
Our measurements confirm that signals in the present realisation travel along hyperbolic geodesics, a~smoking gun signature for the negative curvature of space.
Based on our results, we infer that electric circuit networks could be readily utilized to implement and to experimentally verify the predicted features of the recently studied hyperbolic models of Refs.~\onlinecite{Maciejko:2020,Boettcher:2021,Ikeda:2021,Kollar:2019,Boettcher:2020,Asaduzzaman:2020,Yu:2020,Urwyler:2021}.
We expect the presented methodology for extracting fingerprints of negative curvature to be generalizable to other platforms, in particular to superconducting waveguide resonators that may allow for exciting future incorporation of quantum phenomena~\cite{Kollar:2019}.

\pdfbookmark[0]{Results}{results}\sectitle{Results}

\pdfbookmark[1]{Spectra of Euclidean and hyperbolic drums}{spectra}\subsectitle{Spectra of Euclidean and hyperbolic drums.}
We start by comparing the eigenmodes of Euclidean and hyperbolic drums in the continuum.
The hyperbolic plane, characterized by a~constant negative Gaussian curvature $K < 0$, is naturally embedded in $(2{+}1)$-dimensional Minkowski space as a~hyperboloid sheet with fixed timelike distance from the origin, see \cref{fig:continuum}a.
To solve for the eigenmodes of the wave equation, it is convenient to set $K=-4$ and to employ the stereographic projection (\cref{fig:continuum}a), which maps the hyperbolic plane onto the Poincaré disk, i.e., the unit disk with length element $d s^2 = (1-x^2-y^2)^{-2}(dx^2 + dy^2)$.

The eigenmodes of the hyperbolic drum with $x^2+y^2\leq r_0^2<1$ correspond to the spectrum~\cite{Sarnak:2003,Marklof:2012,Boettcher:2020} of the Laplace-Beltrami operator
\begin{equation}
    \Delta_\mathrm{H} = \left(1-x^2-y^2\right)^2 \Delta_\mathrm{E},
\end{equation}
where $\Delta_\mathrm{E}=(\partial^2/\partial x^2 + \partial^2/\partial y^2)$ is the usual Laplace operator in the Euclidean plane.
Adopting Dirichlet boundary conditions, which yield vanishing amplitude on the disk boundary, the spectrum of the drum is given by solutions~to
\begin{equation}
    -\Delta_\mathrm{g}^{\phantom{n}} u_\mathrm{g}^{n \ell} = \lambda_\mathrm{g}^{n \ell} u_\mathrm{g}^{n \ell}\quad\textrm{with}\quad  \left. u_\mathrm{g}^{n \ell}\right|_{x^2+y^2 = r_0^2} = 0,
    \label{eq:spectrum-eq}
\end{equation}
where $\mathrm{g}\in\{\mathrm{E},\mathrm{H}\}$ indicates the geometry, and $\lambda^{n \ell}_\mathrm{g}$ is the frequency of the mode with angular momentum $\ell$ and with $n$ radial zeroes.
Solutions to \cref{eq:spectrum-eq} are superpositions of Bessel functions (associated Legendre functions) in the Euclidean (hyperbolic) case, cf.~Methods.

\begin{figure}
    \centering
    \includegraphics{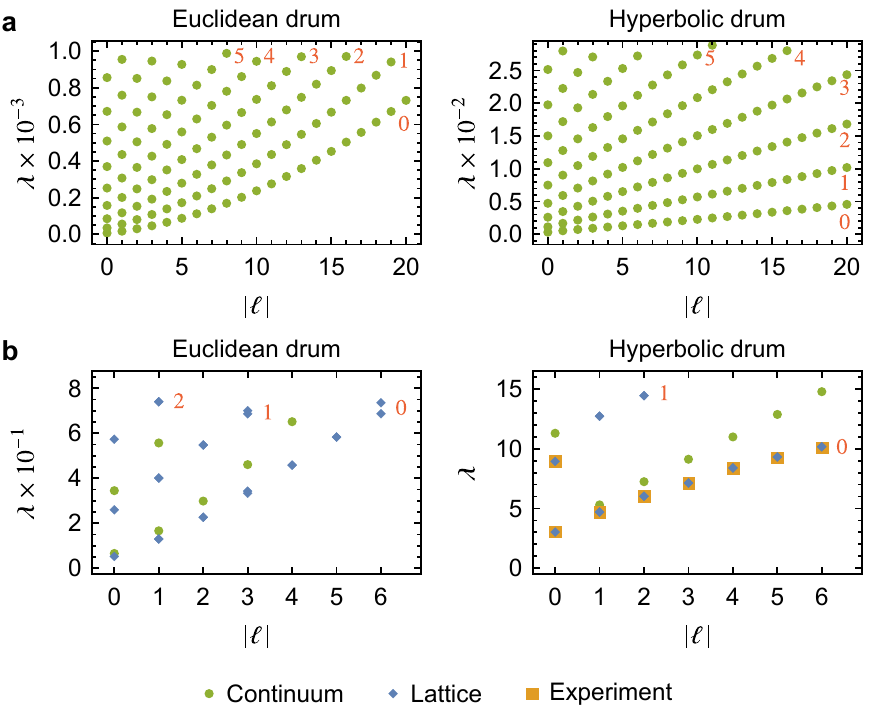}
    \titledcaption{Angular momentum dispersion.}{
        \textbf{a} Rescaled frequency $\lambda_\mathrm{g}^{n \ell}$ vs.\ angular momentum $\ell$ for eigenmodes $u_\mathrm{g}^{n \ell}$ of the continuum Laplace-Beltrami operator, i.e., solutions to \cref{eq:spectrum-eq}, for the Euclidean (left) and hyperbolic (right) geometry.
        For the first six branches, the value of $n$ is indicated by red numbers.
        \textbf{b} Same data for a~Euclidean $\{3,6\}$ (left) and hyperbolic $\{3,7\}$ (right) tessellation, each with $85$ sites. For the hyperbolic lattice, we additionally show the experimental results (orange squares) obtained from the electric circuit.
    }
    \label{fig:dispersion}
\end{figure}

\begin{figure*}[t]
    \centering
    \includegraphics{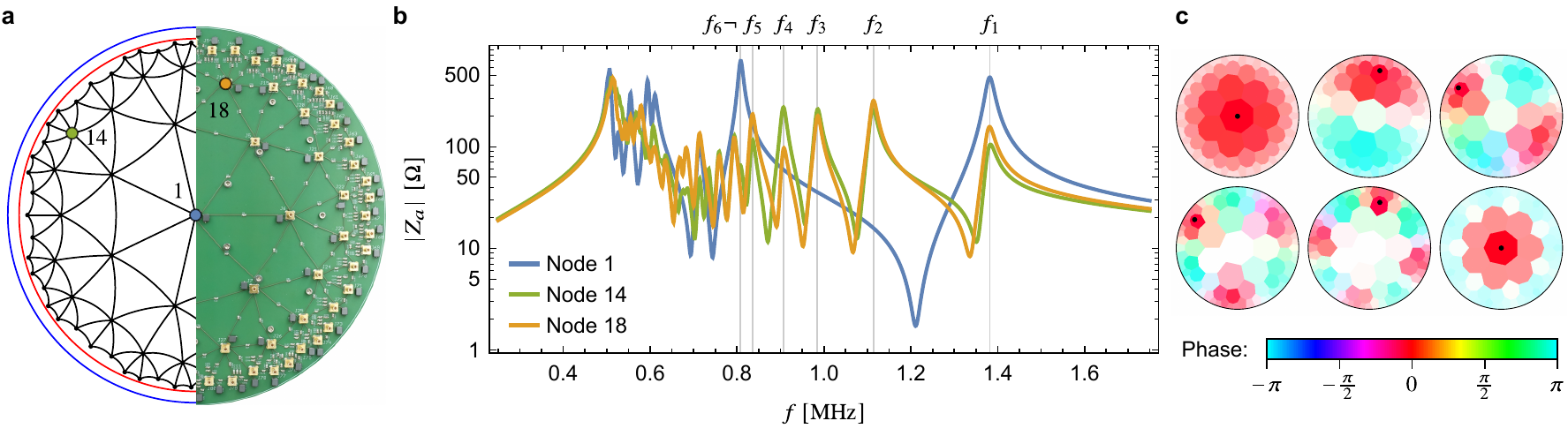}
    \titledcaption{Experimental data.}{
        \textbf{a} Schematic of hyperbolic tessellation (left half) with the unit circle in blue and the circle with radius $r_0=0.94$ in red, and photograph of the electric circuit (right half).
        \textbf{b} Measurement of impedance to ground $Z_a$ of the circuit at node $a$ as a~function of input frequency $f$ for different nodes (see inset legend and panel a~for an identification of the nodes).
        Each impedance peak indicates an eigenmode at that corresponding frequency, which can be excited at the corresponding input node. The highest six frequencies are indicated by vertical grey lines and the corresponding eigenmodes are shown in panel \textbf{c}.
        \textbf{c} Measurement of the voltage profile of the first six eigenmodes (only one mode is shown for each pair of degenerate modes). The saturation encodes the magnitude as a~fraction of the voltage (white denotes $0$ and full saturation $1$) at the input node (black dots), and the color encodes the phase relative to the reference voltage (see legend).
    }
    \label{fig:experiment}
\end{figure*}

We plot in \cref{fig:continuum}b the first few solutions to \cref{eq:spectrum-eq} on the Euclidean vs.\ Poincaré disk for $r_0 = 0.94$, which corresponds to our experimental realisation discussed below.
We observe a~significant reordering of the eigenmodes characterized by $(n,\ell)$: while in the Euclidean case the first eigenmode with $n=1$ is the fourth (not counting degenerate eigenmodes separately), in the hyperbolic case, it is only the sixth mode.
This reordering becomes even more apparent when considering the angular momentum dispersion $\lambda_\mathrm{g}^{n\ell}$ vs.\ $\ell$ displayed in \cref{fig:dispersion}a.
In both the Euclidean and the hyperbolic case, several branches (corresponding to different values of $n$, indicated by red numbers) are discernible.
The spectral reordering manifests as a~reduced slope of the branches (relative to their spacing) compared to their behaviour for the Euclidean drum.
Consequently, eigenmodes with large $\ell$ and small $n$ appear much earlier in the spectrum in hyperbolic compared to Euclidean space.
The spectral reordering is stronger for larger radii $r_0$.
This is intuitively understood from the fact that the circumference of a~hyperbolic drum grows superlinearly with its radius, such that oscillations in the angular direction stretch over larger distances.
This makes them energetically favorable over oscillations in the radial direction, resulting in the observed reordering.

\pdfbookmark[1]{Lattice regularization of the hyperbolic plane}{lattice}\subsectitle{Lattice regularization of the hyperbolic plane.}
To experimentally realise a~hyperbolic drum in an electric circuit network, we discretize the continuous space formed by the hyperbolic plane.
This is achieved by tessellating the hyperbolic plane with regular polygons; a regular tessellation with $q$ copies of $p$-sided polygons meeting at each vertex is conventionally denoted by the Schl\"{a}fli symbol $\{p,q\}$.
The curvature of the continuous space constrains the possible choices of $p$ and $q$: for vanishing curvature (Euclidean plane) they need to satisfy ${(p-2)(q-2)=4}$, while negative curvature (hyperbolic plane) requires ${(p-2)(q-2)>4}$.
A given regular hyperbolic tessellation uniquely fixes the distance between neighboring sites (cf.~Supplementary Note~3), in contrast to the Euclidean case where the distance can be scaled arbitrarily.

Interpreting the vertices as sites of a lattice and the edges as connections between nearest neighbours, we obtain a hyperbolic lattice.
The sites and nearest-neighbour connections form a graph whose Laplacian matrix gives the lattice regularization of the continuum Laplace-Beltrami operator~\cite{Boettcher:2020}, which is fully determined by the topology of the lattice.
The metric of the underlying continuous space is manifested in the connectivity of the lattice sites and therefore in the graph without reference to the positions of the vertices.
However, the positions of the graph nodes (i.e., lattice sites) are relevant for the interpretation of the graph as a lattice when explaining the effective physics.

Different tessellations of the hyperbolic plane are possible, and they generally differ in their symmetries and in how densely their vertices cover the disk.
For our experiments, three different aspects of the modelled lattice are important:
(\emph{i}) the lattice should provide a good approximation of the continuum,
(\emph{ii}) a large fraction of the Poincaré disk should be covered to obtain strong signatures of the negative curvature, and
(\emph{iii}) $\ell=0$ modes should be easy to excite and distinguish from $\ell\neq 0$ modes.
While aspects (\emph{i}) and (\emph{ii}) can both be satisfied by having a sufficiently large number of vertices, in practice, there will be a trade off between the two aspects:
for a~fixed number of vertices, tessellations with larger area per vertex cover a~larger fraction $r_0$ of the Poincaré disk, while for fixed coverage $r_0$ a~good approximation of the continuum is naturally achieved by tessellations that feature small area per vertex (i.e., which tile the hyperbolic plane densely)~\cite{Boettcher:2020}.
Finally, (\emph{iii}) depends on the symmetry properties of the lattice:
a~vertex at the origin of the disk allows for easy excitation and identification of $\ell = 0$ modes and a~high order of rotation symmetry prevents $\ell\neq 0$ modes to have non-vanishing weight at the origin of the disk, which would impede the identification of $\ell=0$ modes.
We analyze and compare several different tessellations with respect to these three aspects in the Supplementary Note~3.
These considerations favour the $\{3,7\}$ tessellation, which exhibits a~seven-fold rotation symmetry with respect to a~site at the centre, and which covers a~disk with radius $r_0=0.94$ with only $85$ sites, see \cref{fig:experiment}a.

\begin{figure*}[t]
    \centering
    \includegraphics{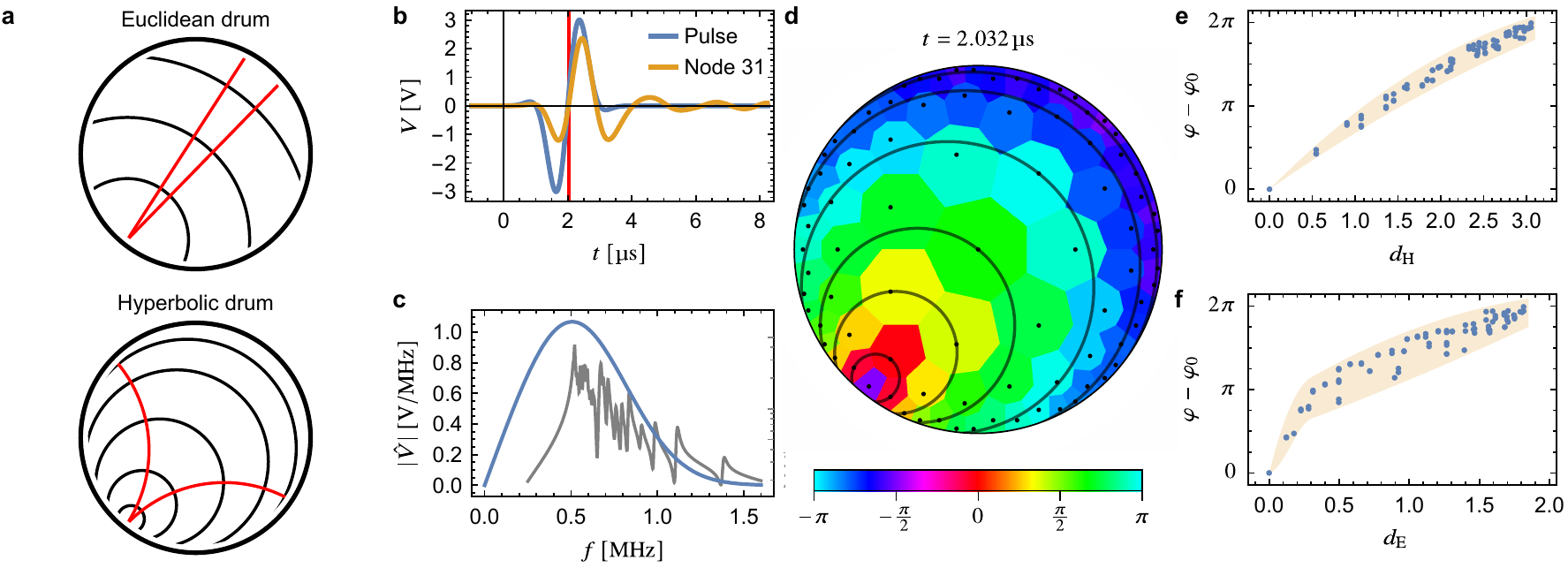}
    \titledcaption{Time-resolved measurement.}{
        \textbf{a} Schematic illustration of the wave propagation after exciting a~Euclidean (top) and hyperbolic (bottom) drum with a~short and spatially localized pulse. The waves travel along geodesics originating from the source (red lines) and wave fronts at different times are given by concentric circles perpendicular to the geodesics. Several equidistant circles with radii $0.5, 1, \dotsc$ (in the appropriate metric) are shown (black circles) for both cases, illustrating distances $d_\mathrm{E}$ and $d_\mathrm{H}$ to the source.
        \textbf{b} Broadband excitation pulse (blue) which is fed as a~current pulse into node $31$ at the boundary, and the voltage response measured at the same node (orange). The time corresponding to the instantaneous phases in panels \textbf{d}--\textbf{f} is marked by a~red vertical line.
        \textbf{c} Frequency spectrum (blue) of the excitation pulse shown in panel \textbf{b}, demonstrating the wide range of frequencies contained in the pulse by comparison to the impedance to ground $\abs{Z_{31}}$ (gray; shown on a logarithmic scale on the right axis from $20$ to $\SI{500}{\ohm}$).
        \textbf{d} Instantaneous phases of the pulse propagating on the hyperbolic drum (see legend) at time $t=\SI{2.032}{\micro\second}$. The nodes are indicated by black dots, and concentric hyperbolic circles with center at node $31$ are shown in black to illustrate the hyperbolic metric.
        \textbf{e} Difference of the instantaneous phase $\varphi$ at each node to the one at the source of the signal (node $31$) $\varphi_0$ vs.\ the hyperbolic distance $d_\mathrm{H}$ to the the source.
        \textbf{f} Difference of the instantaneous phase $\varphi$ at each node to the one at the source of the signal (node $31$) $\varphi_0$ vs.\ the Euclidean distance $d_\mathrm{E}$ to the the source. The shaded region in panels \textbf{e},\textbf{f} indicates the approximate spread of the instantaneous phase as a~function of $d_\mathrm{H}$ and $d_\mathrm{E}$, respectively.
    }
    \label{fig:experiment_pulse}
\end{figure*}

In the long-wavelength-regime, eigenvectors of the Laplacian matrix can be associated with eigenmodes of the Laplace-Beltrami operator in the continuum.
We match them by systematically determining the absolute value of the angular momentum $\ell$ of the eigenvectors by a~Fourier transform of their components on the outermost sites.
Due to the discreteness of the lattice, this analysis is only reliable for modes with sufficiently small $\ell$ and $n$, i.e., in the long-wavelength limit.
Note that while the Laplacian matrix is defined purely on the graph, to define angular momentum we need to interpret the graph as a regular lattice, i.e., identify the vertices with lattice sites.
But, as mentioned above, the (relative) positions of those sites are uniquely defined by the graph via the values of $p$ and $q$.
We extract the angular momentum dispersion for the chosen tessellation, and in \cref{fig:dispersion}b compare it to the corresponding Euclidean $\{3,6\}$ tessellation with the same number of sites.
As in the continuum, a~strong spectral reordering is observed.
This reordering is a~universal feature of the spatial curvature and does, therefore, not rely on the details of the tessellation, as long as it adequately approximates the continuum.

\pdfbookmark[1]{Implementation in an electric circuit}{circuit}\subsectitle{Implementation in an electric circuit.}
In our experiments, the tessellation is realised as an electric circuit network (right half of \cref{fig:experiment}a) with a~node at each site.
Nodes are coupled capacitively among each other and inductively to ground.
The boundary conditions are implemented by additional capacitive coupling of the nodes in the outermost shell to ground.
Effectively, this corresponds to adding one more shell with all nodes shorted to ground, i.e., it represents the lattice equivalent of the Dirichlet boundary conditions.
A~generic electric circuit network is described by Kirchoff's~law
\begin{equation}
    I_a = \sum_b J_{ab}(\omega)V_b,
\end{equation}
where $I_a$ and $V_a$ are the input current and voltage amplitude (for angular frequency $\omega$) at node $a$, respectively.
The matrix $J(\omega)$ is called~\cite{Lee:2018} the grounded circuit Laplacian, and generally depends on $\omega$.
In the continuum limit, the input current $I$ at some position is related to the divergence of the current density $\vec{j}$ via $I=\div\vec{j}$, with $\vec{j}=\sigma\vec{E}=\sigma\grad V$, $\sigma$ the conductivity, $\vec{E}$ the electric field due to an applied voltage $V$, and $\grad$ the del operator (for brevity, we dropped the subscript $\textrm{g}$ indicating the geometry).
Hence, $I = \div\left(\sigma\grad V\right) = \sigma\Delta V$, establishing the interpretation of $J$ as the restriction of the continuum Laplace operator to the grounded circuit.
The impedance to ground of node $a$, $Z_a(\omega)=V_a/I_a$, is fully determined by $J$ and its resonances correspond to eigenmodes of $J$ with eigenvalues $\lambda\propto 1/\omega^2$ (see Methods).
Note that this relationship could be changed to $\lambda\propto \omega^2$ by exchanging the roles of capacitors and inductors in implementing the connections between the nodes resp.~to the ground.

Three types of experiments are performed.
First, an impedance analyzer is used to measure $Z_a$ as a~function of frequency $f=\omega/2\pi$ for each node $a$.
The data for three input nodes are shown in \cref{fig:experiment}b.
Second, these eigenmodes are resonantly excited and their voltage profile is measured using lock-in amplifiers.
For the modes at the highest six frequencies, both magnitude (relative to the voltage at the input node) and phase (relative to a~reference signal) are shown in \cref{fig:experiment}c.
In the final experiment, the circuit is stimulated by the broadband voltage pulse shown in \cref{fig:experiment_pulse}b fed into the circuit as a~current pulse at a~node close to the boundary.
Subsequently, the voltage is measured as a~function of time at each node.
We observe the pulse to propagate in the Poincaré disk (the full time dependence is shown in \phasefront{} and discussed in Supplementary Note~6).
A~snapshot of the instantaneous phase profile (obtained via a Hilbert transform) is shown in \cref{fig:experiment_pulse}d, which visualizes the propagation of the pulse.

\pdfbookmark[1]{Evaluation of the experimental data}{data}\subsectitle{Evaluation of the experimental data.}
We proceed with discussing the results of these three measurements.
Comparing the impedance of input node $1$ (blue curve) to nodes 14 and 18, see \cref{fig:experiment}b, we clearly observe the spectral reordering discussed in the previous section: there are four additional peaks for input node $14$ and $18$ located between the two highest-frequency peaks for input node $1$.
This implies that the second $\ell=0$ mode (i.e.\ the first mode with $n>0$) is the sixth eigenmode.
The explicit values of $\ell$ and $n$ for specific modes can be deduced from the voltage profiles of the eigenmodes obtained in the second experiment, see \cref{fig:experiment}c.
 
We further plot (orange squares in \cref{fig:dispersion}b) the extracted dispersion of the Laplacian frequencies $\lambda_\mathrm{H}^{n,\ell}$ with the angular momentum $\abs{\ell}$, obtained by a~circular Fourier transform of the measured signal.
We observe an almost perfect match with the theoretically predicted values (blue dots in \cref{fig:dispersion}b) for the first few measured modes.
However, higher modes are increasingly difficult to excite and detect, due to the finite resolution in frequency and space.
We remark that the boundary sites of the present experimental realisation of a~hyperbolic lattice could be used to probe holographic dualities.
For each eigenmode of the system, only its angular distribution on the boundary is important (cf.\ the angular momentum dispersion in \cref{fig:dispersion}b), yielding a~novel and universal one-dimensional physical system on the boundary.
We leave a~detailed examination of these intriguing edge modes to future studies.

Finally, we discuss the time-resolved measurements.
We excite the densest region of the frequency spectrum (\cref{fig:experiment}b) using a~current pulse (\cref{fig:experiment_pulse}b) of mean frequency $500$ kHz (\cref{fig:experiment_pulse}c).
By exciting a large number of modes, we approximate the continuum response.
The propagation of the pulse through the circuit network leads to the profile of instantaneous phases depicted in \cref{fig:experiment_pulse}d, where the phase fronts can be easily identified by the positions of equal instantaneous phase.
Since the connectivity of the nodes implements the metric of the Poincaré disk, these phase fronts form concentric hyperbolic circles, highlighted by black circles in \cref{fig:experiment_pulse}d.
This agrees with the theoretical expectation that the signal emanates from the excited node along geodesics, which are the generalization of straight lines in curved space (red lines in \cref{fig:experiment_pulse}a).

Wave fronts are perpendicular to these geodesics and thus constitute concentric circles (black circles in \cref{fig:experiment_pulse}a) up to corrections due to boundary reflections.
In \cref{fig:experiment_pulse}d-f, we have chosen an early time during the excitation such that contributions from such reflections do not have a~significant impact on the measured phases.
Finally, when plotting the phase vs.\ hyperbolic ($d_\mathrm{H}$) and Euclidean ($d_\mathrm{E}$) distance in \cref{fig:experiment_pulse}e and f, respectively, we observe that the correlation of the phase with $d_\mathrm{H}$ is stronger than with $d_\mathrm{E}$.
This manifests that the propagation of the signal indeed follows hyperbolic rather than Euclidean geodesics, thereby verifying that the system realises the hyperbolic rather than Euclidean metric.

\pdfbookmark[0]{Discussion}{discussion}\sectitle{Discussion}

\noindent We have experimentally simulated the negatively curved hyperbolic plane, as evidenced both in the spectral ordering of the Laplace operator and in the signal propagation along curved geodesics. 
With an implementation encompassing only 85 lattice sites, we have readily observed an excellent approximation of the hyperbolic plane; at the same time, no technical constraint hinders significantly enlarging the number of sites in future applications.
In particular, using existing chip manufacturing technology and commercially available components, electric circuits representing lattices with $\sim\!\!10^4$ sites should be within reach.
In combination with the presented results, the efficient fabricability and high scalability of electric circuits elevates them into a versatile platform for emulating classical hyperbolic models, with several advantages over the previously considered methods~\cite{Leonhardt:2006,Kollar:2019}.

First, electric circuits provide easy means for embedding hyperbolic lattices on a flat physical geometry, while allowing for unconnected wire crossings.
Such flexibility could be utilized to include coupling beyond nearest neighbors and to implement the plethora of other hyperbolic tessellations~\cite{Coxeter:1957,Coxeter:1979}.
In particular, going beyond the presented emulation of the Laplace operator in a negatively curved space, the platform allows to emulate much more complex tight-binding models.
These could, for example, be used to test the recently emerging concepts of hyperbolic band theory~\cite{Maciejko:2020,Maciejko:2021,Ikeda:2021}, hyperbolic crystallography~\cite{Boettcher:2021}, and hyperbolic topological insulators~\cite{Yu:2020,Urwyler:2021}.
Electric circuits also excel at providing time- and spatially resolved access to the individual degrees of freedom.

Furthermore, including non-linear and non-reciprocal elements in the network, such as transistors and diodes, is trivial~\cite{Lu:2018}.
This enables experimental investigation of how phenomena like topological insulators \cite{Yu:2020,Urwyler:2021,Ningyuan:2015,Albert:2015}, the non-Hermitian skin effect \cite{Helbig:2020, Lee:2019}, further non-Hermitian topological systems \cite{Gong:2018} or non-linear topological systems \cite{Hadad:2018,Dobrykh:2018,Kotwal:2021} interact with with the negative curvature underlying hyperbolic lattices.
Given their large scalability, electric circuits could be manufactured with the goal to experimentally study non-linear dynamics of systems with sizes that are unwieldy for numerical simulations.
Staying instead within the linear regime, there is a relationship between particles moving freely on geodesics of negatively curved space and deterministic chaos, as illustrated by the Hadamard system~\cite{Balasz:1986}.
In combination with our experimental verification of the signal propagation along the geodesics, this relationship designates electric circuits a~promising experimental platform to investigate classical models of chaos.

Crucially, our work demonstrates the experimental viability of two methods for verifying the hyperbolic nature, i.e., the negative curvature, of the simulated model, which is an important step towards realising more complicated models.
The two methods rely on approximating the Laplace-Beltrami operator using a simple nearest-neighbour tight-binding model and then observing (1) a reordering of eigenmodes compared to flat space, or (2) the propagation of a pulse along hyperbolic geodesics.
These methods are, at least in principle, transferable to other platforms, even though it 
may generally be more challenging to experimentally access the necessary (spatially or time-resolved) information.
However, the first method can be applied in a minimal fashion that requires access to significantly less experimental data.
As we show in \cref{fig:experiment}b, it is sufficient to measure the response (here the impedance to ground) at two vertices, one at the origin and one away from it, in order to distinguish $\ell=0$ from $\ell\neq 0$ modes and observe the predicted mode reordering.
In this respect, note that waveguide resonator circuits, were previously proposed as a platform for realising hyperbolic models as well~\cite{Kollar:2019}.
However, no substantial experimental verification of the curvature has been performed so far.
Our methods could be used to perform a similar analysis on that platform.

Let us finally remark that while coplanar waveguide resonators have been proposed as a promising platform for implementing quantum hyperbolic matter, it is also conceivable~\cite{Lu:2018} that superconducting qubits could potentially be combined with electric circuits in the future.
This suggests another route towards exciting future generalizations of our work to quantum models.
We expect such generalizations to inspire a~new paradigm for designing and measuring holographic toy-models and topological or conformal boundary field theories in discrete geometries. 
In this context, it is worth reminding that theoretical models of hyperbolic quantum systems were proposed~\cite{Zhu:2021}, which still await experimental implementation, including MERA tensor networks\cite{Vidal:2007,Matsueda:2013} and topological quantum memories\cite{Breuckmann:2017,Dennis:2002}.
These efforts have the potential to fundamentally alter our understanding of physics in curved spaces and imply novel views on problems in condensed matter theory, quantum gravity, cosmology, and holography.
\medskip

\let\oldaddcontentsline\addcontentsline     
\renewcommand{\addcontentsline}[3]{}        

{\fontsize{8}{9.6}\selectfont
\vspace{1em}
\pdfbookmark[0]{Methods}{methods}\noindent{\textbf{\large Methods}}\smallskip

\pdfbookmark[1]{Eigenmodes of the Laplace-Beltrami operator}{eigenmodes}\noindent\textbf{Eigenmodes of the Laplace-Beltrami operator.}
The solutions to \cref{eq:spectrum-eq} on the disk $\mathcal{D}_{r_0}$ of radius $r_0<1$ correspond to the eigenmodes of a~drum of radius $r_0$ in the corresponding geometry.
They can be conveniently expressed using special functions.
Going to polar coordinates $(r,\theta)$, one finds (cf.~Supplementary Note~1) for the Euclidean metric
\begin{equation}
    u_\mathrm{E}^{n \ell}(r,\theta) = \mathcal{J}_\ell(k_{n \ell}r)\e^{\i \ell\theta},
\end{equation}
where $\mathcal{J}_\ell(z)$ are the Bessel functions of the first kind and $k_{n \ell}$ is the $(n+1)$-th zero of $k\mapsto\mathcal{J}_\ell(kr_0)$.
From the angular part of the solution it follows that $\ell$ can be interpreted as the angular momentum.
Furthermore, $k_{n \ell} = \frac{z_{\ell,n+1}}{r_0}$, where $z_{\ell,n}$ is the $n$th zero of $\mathcal{J}_\ell(z)$.
The radial zeroes $r_m$ of $u_\mathrm{E}^{n \ell}(r,\theta)$ are then given by
\begin{equation}
    r_m = \frac{z_{\ell,m}}{k_{n \ell}} = r_0\frac{z_{\ell,m}}{z_{\ell,n+1}},
\end{equation}
such that $m=1,2,\dotsc,n$ for the non-trivial zeroes $r_m<r_0$.
Thus, $u_\mathrm{E}^{n \ell}$ has exactly $n$ non-trivial radial zeroes.

For the hyperbolic metric, on the other hand, one finds (cf.~Supplementary Note~1)
\begin{equation}
    u_\mathrm{H}^{n \ell}(r,\theta) = P_{\frac{1}{2}\left(-1+\i k_{n \ell}\right)}^\ell\left(\frac{1+r^2}{1-r^2}\right)\e^{\i \ell\theta}
\end{equation}
with $P_\lambda^\ell(z)$ the associated Legendre functions and $k_{n \ell}$ the $(n+1)$-th zero of $k\mapsto P_{\frac{1}{2}\left(-1+\i k\right)}^\ell\left(\frac{1+r_0^2}{1-r_0^2}\right)$.
Again we can interpret $\ell$ as the angular momentum and $n$ as the number of radial zeroes of $u_\mathrm{H}^{n \ell}$.\medskip

\pdfbookmark[1]{Lattice regularization}{regularization}\noindent\textbf{Lattice regularization.}
The graph Laplacian of a~simple (i.e., undirected) graph is given by
\begin{equation}
    Q = A - D,
\end{equation}
where $D$ is the degree matrix (the diagonal matrix containing the number of adjacent sites for each site as entries) and $A$ the adjacency matrix ($A_{ab}=1$ if sites $a$ and $b$ are adjacent and zero otherwise).
Assuming the graph represents a~lattice regularization of a~continuum, then any function $u(x,y)$ induces a~function on the lattice, via $a\mapsto u(x_a,y_a)=: u_a$, and the action of the Laplacian matrix, $\sum_b Q_{ab}u_b$, can be expressed in terms of the continuum Laplace-Beltrami operator, e.g., following the steps outlined in Ref.~\onlinecite{Boettcher:2020}.

Tessellations of the Euclidean or hyperbolic plane constitute a~lattice regularization of the continuum~\cite{Boettcher:2020}, and the boundaries of the tiles (i.e., vertices and edges) can be interpreted as forming a~graph.
If only a~finite segment of the plane is tiled, the tessellation has a~boundary, which corresponds to vertices of the graph that are attached to fewer edges than the bulk vertices.
Naturally, this is reflected both in the adjacency matrix $A$ as well as in the degree matrix $D$.
However, if we impose Dirichlet boundary conditions for $u(x,y)$ as we do in the main text, then $u$ vanishes on the boundary sites, which allows us to drop them from the matrix description. 
Consequently, we are left only with the bulk part of $Q$.
For a~Euclidean $\{3,6\}$ tessellation, we find (cf.~Supplementary Note~2)
\begin{equation}
    \sum_b Q_{ab}u_b = \frac{3}{2}d^2\Delta_\mathrm{E} u_a + \order{d^3},
    \label{eq:Q_36}
\end{equation}
where $d$ is the distance between sites.
For the hyperbolic tessellation $\{3,7\}$, on the other hand, we find (cf.~Supplementary Note~2)
\begin{equation}
    \sum_b Q_{ab}u_b = \frac{7}{4}h^2\Delta_\mathrm{H} u_a + \order{h^3},
    \label{eq:Q_37}
\end{equation}
where $h=\tanh(d_0)=0.496\,970$, and $d_0=0.545\,275$ is the hyperbolic distance between two neighboring sites in the Poincaré disk representation.
For both tessellations, the leading contribution is the Laplace-Beltrami operator for the appropriate metric, such that eigenstates of $Q$ correspond to $u_\mathrm{g}^{n \ell}$ from \cref{eq:spectrum-eq} and the eigenvalues are proportional to $\lambda_\mathrm{g}^{n \ell}$ (up to higher-order corrections).\medskip

\pdfbookmark[1]{Extraction of angular momentum dispersion}{dispersion}\noindent\textbf{Extraction of angular momentum dispersion.}
The angular momentum dispersion, $\lambda_\mathrm{g}^{n\ell}$ vs.\ $\abs{\ell}$, shown in \cref{fig:dispersion}b is extracted from the spectrum and eigenstates of the graph Laplacian using Fourier analysis on shells of the graph, i.e.\ sites that have approximately the same distance from the disk center and form a~circle.
A~shell can therefore be considered as a~one-dimensional system with periodic boundary conditions with the polar angle taking the role of position.
For each eigenvector $u$, its components on one of the shells, therefore, define a~periodic function $u_\text{shell}(\theta)$ defined at discrete $\theta$.
By first interpolating $u_\text{shell}(\theta)$ and then performing a~discrete Fourier transform on regular samples, we determine the dominant Fourier component which is interpreted as the angular momentum $\abs{\ell}$ of $u$.
For the eigenstates shown in \cref{fig:dispersion}b it is sufficient to consider the outermost shell, but for higher eigenstates, considering additional shells can improve the results.\medskip

\pdfbookmark[1]{Theoretical description of electric circuit}{circuit-theory}\noindent\textbf{Theoretical description of electric circuit.}
In our circuit network, nodes are coupled with capacitance $C$, each node is coupled to ground via an inductance $L$ and the boundary conditions are implemented by adding additional capacitive couplings to ground such that each node is capacitively coupled to seven other components.
The grounded circuit Laplacian is then given by the graph Laplacian $Q$ of the underlying (bulk) lattice and a~contribution from the inductive grounding (neglecting resistances and other parasitic effects):
\begin{equation}
    J(\omega) = -\i\omega C Q+\frac{1}{\i\omega L}\id.
\end{equation}
The spectral decomposition is therefore given by the eigenstates $\psi^\beta$ and eigenvalues $q^\beta$ of the Laplacian matrix, $-Q\psi^\beta=q^\beta \psi^\beta$, with eigenvalues
\begin{equation}
    j^\beta(\omega) = \frac{1-q^\beta\omega^2LC}{\i\omega L}.
\end{equation}
The eigenmode index can be decomposed into the principal and orbital index, $\beta=(n,\ell)$, to match the analytic solution in the continuum.

The inverse of $J$ is called the circuit Green function and can be obtained by expanding $J$ into eigenmodes (here we assume that $J$ is Hermitian and the circuit grounded, as is the case for our circuit) $J(\omega) = \sum_\beta j^\beta(\omega)\psi^\beta\adjo{\psi^\beta}$;
then,
\begin{equation}
    G(\omega) = \sum_\beta \frac{1}{j^\beta(\omega)}\psi^\beta\adjo{\psi^\beta}.
\end{equation}
Assuming current fed into node $a$, i.e., $I_a=\sum_c I\delta_{ca}$, the impedance of that node to ground can be written in terms of the eigenmodes of $J$
\begin{equation}
    Z_a(\omega) = G_{aa}(\omega) = \sum_\beta\frac{1}{j^\beta(\omega)}\abs{\psi^\beta_a}^2,
\end{equation}
and the stationary voltage response, i.e., after equilibration, at some other node $b$ is given by
\begin{equation}
    V_b = G_{ba}(\omega)I_a = \sum_\beta \frac{1}{j^\beta(\omega)}\psi_b^\beta\cconj{\psi^\beta_a}.
\end{equation}
We observe that in both cases the result is a~superposition of eigenmodes of $J$ with the weight proportional to $1/j^\beta(\omega)$, which has a~resonance at
\begin{equation}
    \omega^\beta = \frac{1}{\sqrt{LCq^\beta}}.
\end{equation}
By combining this result with \cref{eq:Q_37} for the bulk-to-lattice correspondence, it follows that a~resonance of $Z_a$ at frequency $f^\beta=\omega^\beta/(2\pi)$ corresponds to an eigenmode of the hyperbolic drum with eigenvalue
\begin{equation}
    \lambda^\beta = \frac{1}{7\pi^2 h^2LC}\frac{1}{(f^\beta)^2}.
    \label{eq:lambda_vs_f}
\end{equation}
This results in a~spectral reversal where the lowest-frequency (small $\lambda$) eigenmodes of the Laplace-Beltrami operator correspond to the highest-frequency (large $f$) oscillations of the electric circuit.
\Cref{eq:lambda_vs_f} is used to plot the experimental data in \cref{fig:dispersion}. 
If the circuit is probed at one of the resonance frequencies, $\omega^\beta$, then the dominant contribution to $V_b$ is
\begin{equation}
    V_b \approx \frac{1}{j^\beta(\omega^\beta)}\psi_b^\beta\cconj{\psi_a^\beta} = \frac{\psi_b^\beta}{\psi_a^\beta}V_a,
\end{equation}
where $j^\beta$ does not diverge in practice due to the presence of small resistive terms (see Supplementary Note~4 for a discussion of the impact of parasitic resistances and Supplementary Note~5 for an extended analysis of measured eigenmodes).
This implies that the voltage profile encodes the eigenvectors~$\psi_b^\beta$.\medskip

\pdfbookmark[1]{Electric circuit parameters}{parameters}\noindent\textbf{Electric circuit parameters.}
The capacitances of the electric circuit are implemented by ceramic capacitors with $C=\SI{1}{\nano\farad}$ and $1\%$ tolerance, the inductances as power inductors with $L=\SI{10}{\micro\henry}$, $5\%$ tolerance and a~minimal quality factor of $40$ at $\SI{1}{\mega\hertz}$.
Nodes on the boundary have additional capacitors $C$ to ground such that each node is connected to seven identical capacitors in total.
Finally, each node is made accessible for in- and ouput via SMB connectors.\medskip

\pdfbookmark[1]{Measurement details}{measurements}\noindent\textbf{Measurement details.}
The impedance measurements were performed in a~two-terminal measurement configuration using a~Zurich Instruments MFIA $\SI{5}{\mega\hertz}$ impedance analyzer.
A~short/open compensation routine was used to remove the residual impedance and stray capacitance of the test fixture.
The impedance of all $85$ circuit nodes has been recorded for frequencies in the range from $\SI{250}{\kilo\hertz}$ to $\SI{1.75}{\mega\hertz}$.
To exclude transmission line effects in the measurement, the maximum cable length was restricted to be below $\SI{1.8}{\metre}$.

For the measurement of the voltage profiles of the eigenmodes, a~reference voltage signal and phase sensitive detection is needed.
This was achieved by using three Zurich Instrument MFIA $\SI{5}{\mega\hertz}$ impedance analyzers as lock-in amplifiers synchronized in frequency and phase.
Each mode was excited by a~current signal of the corresponding frequency fed into the node with the highest impedance peak at that frequency.
The current signal was obtained by applying the sinusoidal reference voltage signal with fixed peak-to-peak voltage of $\SI{1}{\volt}$ produced by one of the lock-in amplifiers to a~shunt resistor of $\SI{12}{\ohm}$.
The other two lock-in amplifiers were used to measure the voltages of the different nodes.
All voltage signals demodulated with the reference signal were filtered with a~digital low-pass filter of eighth order and a~cutoff frequency of $f_{-3\nunit{dB}}\,=\,\SI{0.7829}{\hertz}$.
The readout of the real and imaginary part of the voltage took place after at least $16$ filter time constants which corresponds to at least $99\,\%$ settling of the low-pass filters in a~step response.

The time-resolved measurements were carried out by seven Picoscope 4824, which are eight channel USB oscilloscopes with $\SI{20}{\mega\hertz}$ bandwidth and $\SI{12}{\bit}$ resolution.
In the experiment, the circuit was stimulated at node $31$ by the broadband pulse
\begin{equation}
    V(t) = V_0\sin(2\pi ft)\e^{-\frac{1}{2}\left(4\left(ft-1\right)\right)^2}
\end{equation}
with $V_0=\SI{4.3}{\volt}$ and $f=\SI{500}{\kilo\hertz}$.
The pulse is generated by a~$\SI{50}{\ohm}$ function generator and the output current was fed directly into the input node.
Since the oscilloscopes do not provide a~separate trigger channel, one channel of each instrument was fed with a~rectangular pulse synchronized with the excitation pulse to trigger the instruments.
They used an edge trigger at $\SI{1}{\volt}$ in rapid trigger mode and sampled with $40\nunit{MS/s}$, i.e., every $\SI{25}{\nano\second}$.
Assuming equal behaviour of the circuit under repeated stimulation, which was verified during the measurement process by repeating the process described below ten times, the measurement was performed in two steps.
First, the seven oscilloscopes were used to measure the voltage at nodes $1$ through $49$ (including the input node $31$), then, in the second run, the input node and nodes $38$ through $85$ were measured.
Finally, the measured real-valued signals $V(t)$ were transformed into complex-valued ones using the Hilbert transform, therefore giving access to the instantaneous phase as the argument of the complex-valued signal
\begin{equation}
    v(t) = V(t) + \frac{\i}{\pi}\int_{-\infty}^\infty d{\tau}\frac{V(\tau)}{t-\tau}.
\end{equation}

\medskip

\pdfbookmark[0]{Data availability}{data-availability}\sectitle{Data availability}

\noindent All the data (both experimental data and data obtained numerically) used to arrive at the conclusions presented in this work are publicly available in the following data repository: \href{https://doi.org/10.3929/ethz-b-000503548}{https://doi.org/10.3929/ethz-b-000503548}.

\medskip

\pdfbookmark[0]{Code availability}{code-availability}\sectitle{Code availability}

\noindent All the Wolfram Language code used to generate and/or analyze the data and arrive at the conclusions presented in this work is publicly available in the form of annotated Mathematica notebooks in the following data repository: \href{https://doi.org/10.3929/ethz-b-000503548}{https://doi.org/10.3929/ethz-b-000503548}.
}

\pdfbookmark[0]{References}{references}
\putbib
\end{bibunit}

{\fontsize{8}{9.6}\selectfont

\medskip

\pdfbookmark[0]{Acknowledgements}{acknowledgements}\sectitle{Acknowledgements}

\noindent P.~M.~L. and T.~B. were supported by the Ambizione grant No.~185806 by the Swiss National Science Foundation.
T.~N. acknowledges support from the European Research Council (ERC) under the European Union’s Horizon 2020 research and innovation programm (ERC-StG-Neupert-757867-PARATOP).
The work in W\"urzburg is funded by the Deutsche Forschungsgemeinschaft (DFG, German Research Foundation) through Project-ID 258499086 - SFB 1170 and through the W\"urzburg-Dresden Cluster of Excellence on Complexity and Topology in Quantum Matter -- \textit{ct.qmat} Project-ID 39085490 - EXC 2147.
T.He. was  supported by a~Ph.D. scholarship of the Studienstiftung des deutschen Volkes.
I.~B. acknowledges support from the University of Alberta
startup fund UOFAB Startup Boettcher and Natural Sciences and Engineering Research Council of Canada (NSERC) Discovery Grants RGPIN-2021-02534 and DGECR-2021-00043.

\medskip

\pdfbookmark[0]{Author contributions}{contributions}\sectitle{Author contributions}

\noindent R.T.~initiated the project, and together with T.N.~and T.B.~led the collaboration. P.M.L., A.S., L.K.U., T.Ho., T.He.~and T.B.~performed the theoretical analysis of the hyperbolic tessellations. P.M.L., A.S.~and A.V.~designed the electric circuit. S.I., H.B.~and T.K.~carried out the measurements, and together with P.M.L.~and A.S.~analyzed the collected data. P.M.L., A.S., I.B., T.N.~and T.B.~wrote the manuscript. 
P.M.L., A.S., L.K.U., T.Ho., T.He., A.V., M.G., C.H.L., S.I., H.B., T.K., I.B., T.N., R.T., and T.B. discussed together and commented on the manuscript. 

\medskip

\pdfbookmark[0]{Competing interests}{interests}\sectitle{Competing interests}

\noindent The authors declare no competing interests.

\medskip

\pdfbookmark[0]{Additional information}{additional}\sectitle{Additional information}

\noindent\textbf{Supplementary information.} The online version of the manuscript is accompanied with supplementary materials, which include: Supplementary Notes~1 to~6, Supplementary Figures~1 to~8, and Supplementary Movie~1.

\smallskip

\noindent\textbf{Correspondence} and requests for materials should be addressed to T.~Neupert, R.~Thomale, and T.~Bzdu\v{s}ek.
}

\let\addcontentsline\oldaddcontentsline     
\clearpage

\begin{bibunit}
\onecolumngrid
\stepcounter{myfigure}
\renewcommand\thesection{\arabic{section}}
\renewcommand\thesubsection{\alph{subsection}}
\makeatletter
\renewcommand{\fnum@figure}{\textbf{Supplementary Figure~\thefigure}}
\makeatother

\titleformat{\section}
  {\centering\normalfont\fontsize{10pt}{11pt}\selectfont\bfseries\selectfont\uppercase}{\uppercase{Supplementary Note~\thesection.}}{1em}{}

\setcounter{page}{1}
\setcounter{equation}{0}
\setcounter{section}{0}
\setcounter{figure}{0}

\title{Supplementary Information\texorpdfstring{ to:\smallskip\\Simulating hyperbolic space on a circuit board}{}}

\author{Patrick M. Lenggenhager\,\orcidlink{0000-0001-6746-1387}}\thanks{\eqcontrib{}}
\affiliation{Condensed Matter Theory Group, Paul Scherrer Institute, 5232 Villigen PSI, Switzerland}
\affiliation{Department of Physics, University of Zurich, Winterthurerstrasse 190, 8057 Zurich, Switzerland}
\affiliation{Institute for Theoretical Physics, ETH Zurich, 8093 Zurich, Switzerland}

\author{Alexander Stegmaier\,\orcidlink{0000-0002-8864-5182}}\thanks{\eqcontrib{}}
\affiliation{Institut für Theoretische Physik und Astrophysik, Universität Würzburg, 97074 Würzburg, Germany}

\author{Lavi K. Upreti\,\orcidlink{0000-0002-1722-484X}}
\affiliation{Institut für Theoretische Physik und Astrophysik, Universität Würzburg, 97074 Würzburg, Germany}

\author{Tobias Hofmann\,\orcidlink{0000-0002-1888-9464}}
\affiliation{Institut für Theoretische Physik und Astrophysik, Universität Würzburg, 97074 Würzburg, Germany}

\author{Tobias Helbig\,\orcidlink{0000-0003-1894-0183}}
\affiliation{Institut für Theoretische Physik und Astrophysik, Universität Würzburg, 97074 Würzburg, Germany}

\author{Achim Vollhardt\,\orcidlink{0000-0002-4424-1127}}
\affiliation{Department of Physics, University of Zurich, Winterthurerstrasse 190, 8057 Zurich, Switzerland}

\author{Martin Greiter}
\affiliation{Institut für Theoretische Physik und Astrophysik, Universität Würzburg, 97074 Würzburg, Germany}

\author{Ching Hua Lee}
\affiliation{Department of Physics, National University of Singapore, Singapore 117551, Republic of Singapore}

\author{Stefan Imhof}
\affiliation{Physikalisches Institut, Universität Würzburg, 97074 Würzburg, Germany}

\author{Hauke Brand\,\orcidlink{0000-0003-4209-2663}}
\affiliation{Physikalisches Institut, Universität Würzburg, 97074 Würzburg, Germany}

\author{Tobias Kie\ss ling}
\affiliation{Physikalisches Institut, Universität Würzburg, 97074 Würzburg, Germany}

\author{Igor Boettcher\,\orcidlink{0000-0002-1634-4022}}
\affiliation{Department of Physics, University of Alberta, Edmonton, Alberta T6G 2E1, Canada}
\affiliation{Theoretical Physics Institute, University of Alberta, Edmonton, Alberta T6G 2E1, Canada}

\author{Titus Neupert\,\orcidlink{0000-0003-0604-041X}}\thanks{\comms{}}
\affiliation{Department of Physics, University of Zurich, Winterthurerstrasse 190, 8057 Zurich, Switzerland}

\author{Ronny Thomale\,\orcidlink{0000-0002-3979-8836}}\thanks{\comms{}}
\affiliation{Institut für Theoretische Physik und Astrophysik, Universität Würzburg, 97074 Würzburg, Germany}

\author{Tom\'{a}\v{s} Bzdu\v{s}ek\,\orcidlink{0000-0001-6904-5264}}\thanks{\comms{}}
\affiliation{Condensed Matter Theory Group, Paul Scherrer Institute, 5232 Villigen PSI, Switzerland}
\affiliation{Department of Physics, University of Zurich, Winterthurerstrasse 190, 8057 Zurich, Switzerland}

\maketitle

\newpage
\onecolumngrid

%
\renewcommand{\tocname}{List of Supplementary Notes}
\tableofcontents

\section{Eigenmodes of the Laplace-Beltrami operator}\label{sec:continuum_eigenmodes}
The general Laplace-Beltrami operator for a~given metric tensor $g_{ij}$ is
\begin{equation}
	\Delta_g = \frac{1}{\sqrt{\det(g)}}\partial_i\left(\sqrt{\det(g)}g^{ij}\partial_j\right),
\end{equation}
where $g^{ij}$ is the matrix inverse of $g_{ij}$.
In Euclidean space, we have $(g_\mathrm{E})_{ij}=\delta_{ij}$, such that $\Delta_\mathrm{E}=\partial_x^2+\partial_y^2$, the usual Laplace operator.
In contrast, for the Poincaré disk representation of the hyperbolic plane ($\mathcal{D}=\{(x,y)\in\mathbb{R}^2\,|\,x^2+y^2=r^2<1\}$ with length element $\dd s^2 = (1-x^2-y^2)^{-2}(dx^2 + dy^2)$ corresponding to constant negative curvature $K=-4$),
\begin{equation}
	(g_\mathrm{H})_{ij} = (1-r^2)^{-2}\delta_{ij},
\end{equation}
such that the Laplace-Beltrami operator is
\begin{equation}
	\Delta_\mathrm{H} = \left(1-\left(x^2+y^2\right)\right)^2\left(\partial_x^2+\partial_y^2\right).
\end{equation}

We now consider the disk $\mathcal{D}_{r_0}:=\{(x,y)\in\mathbb{R}^2\left|x^2+y^2\leq r_0^2\right.\}\subset \mathcal{D}$ and rewrite the Laplace-Beltrami operator in polar coordinates $x=r\cos(\theta)$, $y=r\sin(\theta)$:
\begin{align}
	\Delta_\mathrm{E} &= \partial_r^2 + \frac{1}{r}\partial_r+\frac{1}{r^2}\partial_\theta^2,\\
	\Delta_\mathrm{H} &= \left(1-r^2\right)^2\left(\partial_r^2 + \frac{1}{r}\partial_r+\frac{1}{r^2}\partial_\theta^2\right).
\end{align}
We are interested in eigenmodes of $-\Delta_\mathrm{g}$, where $\mathrm{g}\in\{\mathrm{E},\mathrm{H}\}$ indicates the geometry, i.e.\ solutions to the Dirichlet problem
\begin{equation}
	(\Delta_\mathrm{g}+\lambda)u(x,y) = 0,\qquad \left.u(x,y)\right|_{(x,y)\in\partial\mathcal{D}_{r_0}}=0.
	\label{SI:eq:Dirichlet}
\end{equation}

\subsection{Euclidean space}
We first discuss the solutions to Supplementary Equation~(\ref{SI:eq:Dirichlet}) in the Euclidean case.
The differential equation is separable, such that we can make the ansatz $u(x,y)=R(r)\Theta(\theta)$ and find
\begin{equation}
	-\frac{\Theta''(\theta)}{\Theta(\theta)} = \frac{r^2R''(r) + rR'(r) + r^2\lambda R(r)}{R(r)}.
\end{equation}
Since, we are on the disk, $\Theta(\theta+2\pi)=\Theta(\theta)$, such that $\Theta(\theta)=\e^{\i \ell\theta}$ for $\ell\in\mathbb{Z}$ and
\begin{equation}
	r^2R''(r)+rR'(r)+(k^2r^2-\ell^2)R(r) = 0,
\end{equation}
where we substituted $\lambda=k^2$.
With the further substitution $\rho=kr$, we obtain
\begin{equation}
	\rho^2R''(\rho)+\rho R'(\rho)+(\rho^2-\ell^2)R(\rho) = 0,
\end{equation}
which is the Bessel equation, such that the solutions are given by the Bessel functions of the first kind
\begin{equation}
	u_\mathrm{E}^{n\ell}(x,y) = \mathcal{J}_\ell(k_nr)\e^{\i \ell\theta}
\end{equation}
where $k_n=z_n/r_0$ and $z_n$ is the $n$-th root of $\mathcal{J}_\ell$.

\subsection{Hyperbolic space}
We proceed analogously in the hyperbolic case, where the same ansatz $u(x,y)=R(r)\Theta(\theta)$ results in
\begin{equation}
	(1-r^2)^2r^2R''(r) + (1-r^2)^2rR'(r) + (\lambda r^2 - \ell^2(1-r^2)^2)R(r) = 0.
\end{equation}
Introducing $s := (1+r^2)/(1-r^2)$, this can be rewritten as
\begin{equation}
    2sR'(s) - 4r^2(1-s^2)R''(s)+(\lambda r^2 - \ell^2(1-r^2)^2)R(s) = 0
\end{equation}
Dividing by $-4r^2$ and setting $\lambda=-4q(q+1)=1+k^2$, we find
\begin{equation}
	\left((1-s^2)\partial_s^2 - 2s\partial_s + \left(q(q+1) - \ell^2\frac{1}{1-s^2}\right)\right)R(s) = 0,
\end{equation}
whose solutions are the associated Legendre functions $P_q^\ell(s)$,
such that we obtain
\begin{equation}
	u_{n\ell}(x,y) = P_{\frac{1}{2}\left(-1+\i k_{n\ell}\right)}^\ell\left(\frac{1+r^2}{1-r^2}\right)\e^{\i \ell\theta}
	\label{SI:eq:cont_eigenmode}
\end{equation}
with $k_{n\ell}$ being the $n$-th root of
\begin{equation}
	k\mapsto P_{\frac{1}{2}\left(-1+\i k\right)}^\ell\left(\frac{1+r_0^2}{1-r_0^2}\right)
\end{equation}
and $\ell\in\mathbb{Z}$ as in the Euclidean case.

\section{Lattice regularization of the graph Laplacian}\label{sec:lattice_regularization}
As discussed in the Methods, the graph Laplacian can be approximated by the continuum Laplace-Beltrami operator in the leading order in the distance between lattice sites, see e.g., 
Eqs.~(8) and~(9) in Methods.
In Supplementary Notes~\ref{sec:lattice_regularization:Euclidean} and~\ref{sec:lattice_regularization:hyperbolic} we present a~detailed derivation of this expansion for regular tessellations with equivalent sites where all distances between adjacent sites are equal (in the corresponding metric).
Such tessellations are called Archimedian.
They are generally denoted by their vertex configuration {$n_1.n_2.\cdots.n_q$}, where $n_1,n_2,\dotsc,n_q$ give the number of sides of the $q$ regular polygons meeting at each vertex.
The tessellations considered in the main text are a~special case called Platonic tessellations, because they have $q$ copies of the same regular $p$-gon meeting at each vertex.
In Supplementary Note~\ref{sec:lattice_regularization:cont_approx} we briefly discuss how to quantify how well a certain tessellation approximates the continuum.

We now consider Archimedian tessellations of the unit disk for both Euclidean and hyperbolic space (in the Poincaré disk representation) with Dirichlet boundary conditions imposed.
For convenience, we parametrize the coordinates $(x,y)$ of the Euclidean plane and the Poincaré disk using complex numbers $z:=x+\i y$, where $z$ lies in the infinite complex plane for the Euclidean and in the complex unit disk for the hyperbolic case.
The boundary condition implies that all vertices, including those on the non-vanishing boundary, are equivalent.
Recall that the graph Laplacian is a~matrix $Q=A-D$ with entries $Q_{ab}$ and any test function $u(z)$ on the complex unit disk induces a~function on the lattice, via $a\mapsto u(z_a)=u_a$.
We closely follow Appendix B of Ref.~\onlinecite{Boettcher:2020} to express the action of $Q$ on $u_a$ in terms of the Laplace-Beltrami operator.
The action of the graph Laplacian $Q$ on the test function $a\mapsto u_a$ at an arbitrary site $a$ (using Einstein's summation convention) is then
\begin{equation}
    Q_{ab}u_b = A_{ab}u_b - D_{ab}u_b = \sum_{i=1}^q u(z_{a+e_k}) - qu(z_a),
    \label{SI:eq:graph_Laplacian}
\end{equation}
where $z_{a+e_i}$ denote the position of the sites adjacent to site $a$.

\subsection{Euclidean space}\label{sec:lattice_regularization:Euclidean}
We first discuss a~Euclidean tessellation with coordination $q$, i.e., where each site has $q$ adjacent sites.
Let $d$ be the distance between two adjacent sites, then
\begin{equation}
    z_{a+e_i} = z_a+d\e^{\i\phi_a}\e^{\i\frac{2\pi}{q}(i-1)} =: z_a + dw_{ai},
\end{equation}
where $\phi_a$ is a~site-dependent phase factor, and we can expand $u(z_{a+e_i})$ in powers of $d$:
\begin{equation}
    \begin{split}
        u(z_{a+e_i}) &= u(z_a) + \left.\dv{}{d}u(z_{a+e_i})\right|_{d=0}d + \frac{1}{2}\left.\dv[2]{}{d}u(z_{a+e_i})\right|_{d=0}d^2 + \order{d^3}\\
        &= u(z_a) + \left.\left(w_{ai}\partial_z+\bar{w}_{ai}\bar{\partial}_z\right)u(z)\right|_{z=z_a}d + \frac{1}{2}\left.\left(w_{ai}\partial_z+\bar{w}_{ai}\bar{\partial}_z\right)^2u(z)\right|_{z=z_a}d^2 + \order{d^3}
    \end{split}    
\end{equation}
with $\partial_z=\partial/\partial_z$ and $\bar{\partial}_z=\partial/\partial_{\bar{z}}$ and $\bar{\cdot}$ denoting complex conjugation.
Note that for any $m\in\mathbb{Z}$
\begin{equation}
    \sum_{i=1}^qw_{ai}^m = \e^{\i\phi_a}\sum_{i=1}^q\e^{\i\frac{2\pi m}{q}(i-1)} = 0,
\end{equation}
and $\abs{w_{ai}}=1$, such that $\phi_a$ drops from the subsequent calculations:
\begin{equation}
    \sum_{i=1}^qu(z_{a+e_a}) = qu(z_a) + qd^2\left.\partial_z\bar{\partial}_zu(z)\right|_{z=z_a} + \order{d^3}.
\end{equation}
Since $\Delta_\mathrm{E} = 4\partial_z\bar{\partial}_z$, we finally find
\begin{equation}
    Q_{ab}u_b = \frac{q}{4}d^2\Delta_\mathrm{E}u(z_a) + \order{d^3}.
\end{equation}
With $q=6$ for a~$\{3,6\}$ tessellation this reproduces Eq.~(8) in Methods. 

\subsection{Hyperbolic space}\label{sec:lattice_regularization:hyperbolic}
We proceed analogously for hyperbolic tessellations with coordination $q$ and hyperbolic distance $d_0$ between adjacent sites.
Here it is helpful to first transform the Poincaré disk by the automorphism
\begin{equation}
    z\mapsto v(z) = \frac{z_a-z}{1-z\bar{z}_a}.
\end{equation}
This transformation corresponds to a~$\pi$-rotation that exchanges $z_a$ and the origin. In particular, note that it squares to identity, implying that $z\mapsto v(z)$ and its inverse $z\mapsto v^{-1}(z)$ are equivalent.
Recall further that the hyperbolic distance between the origin and an arbitrary point $z$ in the unit disk takes the form $d=\arctanh(\abs{z})$.
In the transformed coordinates, $z_{a+e_i}$ takes the simple form
\begin{equation}
    v_{a+e_i} = v(z_{a+e_i}) = h\e^{\i\phi_a}\e^{\i\frac{2\pi}{q}(i-1)} = hw_{ai}
\end{equation}
with $h=\tanh(d_0)$, $u(z_{a+e_i}) = u(z(v_{a+e_i}))$, and $\phi_a$ being again a~site-dependent phase factor that subsequently drops out from the calculations.
Expanding in powers of $h$, we obtain
\begin{equation}
    \begin{split}
        u(z_{a+e_i}) &= u(z_a) + \left.\dv{}{h}u(z(v_{a+e_i}))\right|_{h=0}h + \frac{1}{2}\left.\dv[2]{}{h}u(z(v_{a+e_i}))\right|_{h=0}h^2 + \order{h^3}\\
        &= u(z_a) -(1-\abs{z_a}^2)\left.\left(w_{ai}\partial_z+\bar{w}_{ai}\bar{\partial}_z\right)u(z)\right|_{z=z_a}h + \frac{1}{2}(1-\abs{z_a}^2)^2\left.\left(w_{ai}\partial_z+\bar{w}_{ai}\bar{\partial}_z\right)^2u(z)\right|_{z=z_a}h^2 + \order{h^3}
    \end{split}
\end{equation}
Since $w_{ai}$ are still the same as in the Euclidean case, Supplementary Equation~(\ref{SI:eq:graph_Laplacian}) becomes
\begin{equation}
    Q_{ab}u_b = \frac{q}{4}h^2\left.(1-\abs{z}^2)^2\Delta_\mathrm{E}u(z)\right|_{z=z_a} + \order{h^3}
    \label{SI:eq:graph_Laplacian_expansion}
\end{equation}
and recalling that $(1-\abs{z}^2)^2\Delta_\mathrm{E}=\Delta_\mathrm{H}$, we finally arrive at
\begin{equation}
    Q_{ab}u_b = \frac{q}{4}h^2\Delta_\mathrm{H}u(z_a) + \order{h^3}.
\end{equation}
With $q=7$ for a~$\{3,7\}$ tessellation this reproduces Eq.~(9) in Methods. 

\subsection{Approximating the continuum}\label{sec:lattice_regularization:cont_approx}
How faithfully a given tessellation approximates the continuum with respect to the Laplace-Beltrami operator can be quantified according to several different aspects.
Recall that, according to Supplementary Equation~(\ref{SI:eq:graph_Laplacian_expansion}), the graph Laplacian can be interpreted as the leading-order term of an expansion in $h=\tanh(d_0)$ (where $d_0$ is the hyperbolic distance between neighboring sites) of the Laplace-Beltrami operator.
While this allows us to compare different tessellations, it does not directly quantify how good the approximation is for any particular tessellation.
To perform such a quantitative assessment, certain properties can be computed both on a continuous disk as well as on the lattice and then compared to each other.

\begin{figure}
	\centering
	\includegraphics{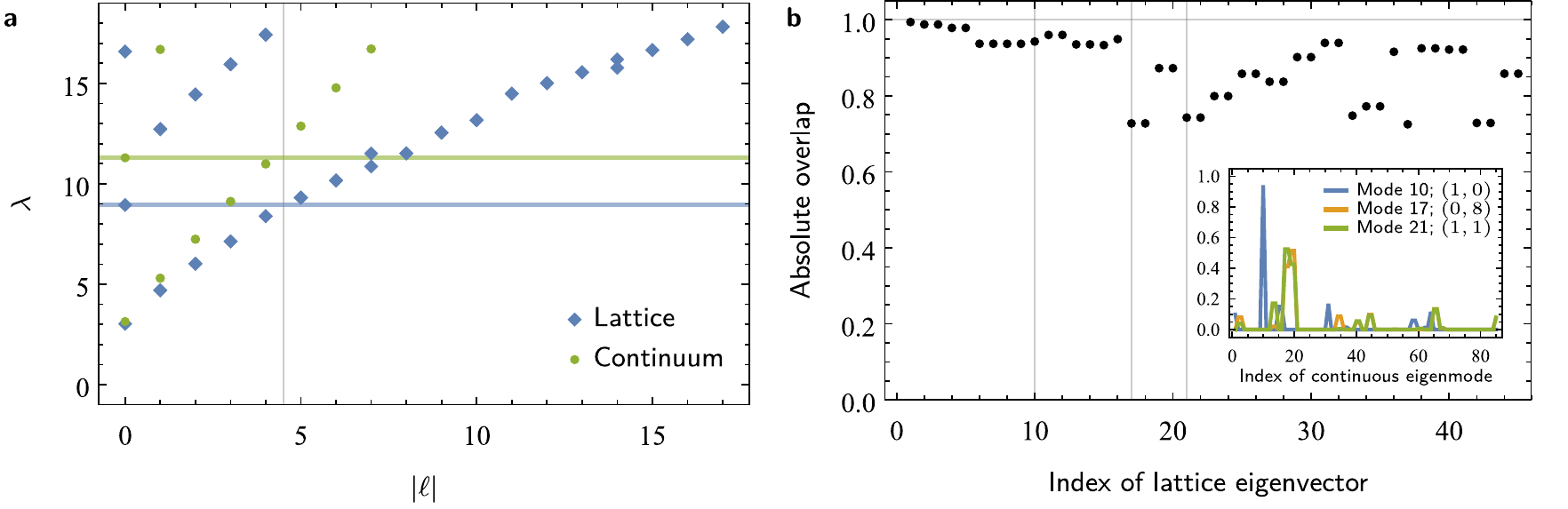}
	\titledcaption{Comparison between lattice and continuum.}{
	    \textbf{a} Eigenvalues $\lambda$ of the Laplace-Beltrami operator (green disks) with Dirichlet boundary conditions at $r_0=0.94$ and the graph Laplacian (blue diamonds) of the graph obtained from the $\{3,7\}$ tessellation with $85$ sites as a function of the absolute value of the angular momentum quantum number $\ell$. Horizontal lines indicate the value of $\lambda$ for the $(1,0)$ mode emphasizing the reordering of the modes compared to flat space (see \cref{fig:dispersion} in the main text).
	    \textbf{b} Absolute value of the overlap of the eigenvectors of the graph Laplacian (with index according to increasing eigenvalue given on the horizontal axis) with the discretized eigenmodes of the Laplace-Beltrami operator for the first $45$ eigenvectors.
	    The inset shows examples for the overlap of three eigenvectors (indicated by vertical gray lines in the main plot) with the first $85$ eigenmodes of the Laplace-Beltrami operator.
	    For each, the maximum overlap is identified and from the corresponding eigenmode $n$ and $\ell$ are extracted (see legend).
	    Note that in the panel the total overlap of the graph Laplacian's eigenvectors with the corresponding eigenmodes $(n,\pm\abs{\ell})$ is given (where the total is defined as the square root of the sum of squares of the individual overlaps), while in the inset the overlap with $(n,+\abs{\ell})$ and $(n,+\abs{\ell})$ is given separately.
	}
	\label{SI:fig:cont_approx}
\end{figure}

For example, in Ref.~\onlinecite{Boettcher:2020} the authors compute the ground state energy and Green function of the Hamiltonian given by $-A$, where $A$ is the adjacency matrix of the graph induced by the tessellation.
In the main text we have compared the ordering of the eigenmodes of the Laplace-Beltrami operator (with appropriate boundary conditions) according to increasing eigenvalues to the one of the graph Laplacian, see also Supplementary Figure~\ref{SI:fig:cont_approx}a.
To formulate a more quantitative criterion, we consider the eigenmodes directly, computing the overlap of the eigenvectors of the graph Laplacian with the discretized eigenmodes of the Laplace-Beltrami operator, shown in Supplementary Figure~\ref{SI:fig:cont_approx}b.
In addition, the Laplacian eigenvalues can also be quantitatively compared; we do the latter in Supplementary Note~\ref{sec:approx_continuum}, when comparing different tessellations.

For both comparisons, the first step is to match the eigenvectors to appropriate eigenmodes.
This is achieved by first computing the overlap of a given eigenvector of the graph Laplacian with the $85$ eigenmodes of the Laplace-Beltrami operator with lowest eigenvalue, and by subsequently determining the quantum numbers $n$ and $\abs{\ell}$ (see Supplementary Note~\ref{sec:continuum_eigenmodes}) of the modes with largest overlap (see inset of Supplementary Figure~\ref{SI:fig:cont_approx}b).
Here, by overlap we mean the dot product of normalized eigenmodes for the graph vs. continuum Laplacian.
We observe in Supplementary Figure~\ref{SI:fig:cont_approx}b that the maximal overlap is very close to $1$ up to (and excluding) mode $17$, which corresponds to the first mode whose order does not agree with the continuum case anymore: mode $17$ in the continuum is the one with $(n,\abs{\ell})=(1,1)$, while on the lattice it is $(0,8)$.
For the mode reordering to be observable, the overlap has to be close to $1$ for all the modes up to (and including) the $(1,0)$ mode, which in our case is mode $10$.
Note that due to the small number of lattice sites, the quantitative agreement of the eigenvalues is not particularly good, but is is improved significantly when increasing the number of sites, cf.~Supplementary Figure~\ref{SI:fig:tessellations:fixed_r0}.
However, for our purposes the overlap of the eigenmodes is sufficient to guarantee the reordering.

\section{Comparison of tessellations of hyperbolic space}\label{sec:tessellations}
In Supplementary Note~\ref{sec:lattice_regularization} we have derived that corrections of the graph Laplacian to the continuum Laplace-Beltrami operator are of third order in $h=\tanh(d_0)$ with $d_0$ being the hyperbolic distance between adjacent lattice sites.
This agrees with the intuition that the density of the tessellation determines the accuracy of the approximation of the continuum.
This fact should not be misunderstood, however, as implying that the different tessellations differ only in the positions of the sites.
On the contrary, different Archimedean tessellations, each specified by $n_1.n_2.\cdots.n_q$, where $q$ is the number of polygons joining at each site and $n_i$ the number of sides of the $i^\textrm{th}$ polygon, differ even if viewed as graphs.

For any planar graph, we can define the Euler characteristic per vertex
\begin{equation}
    \Delta\chi = \Delta V - \Delta E + \Delta F,
\end{equation}
where $\Delta V = 1$ is the number of vertices per vertex, $\Delta E$, the number of edges per vertex, and $\Delta F$ the number of faces per vertex.
The graph induced by the Archimedean tessellation $n_1.n_2.\cdots.n_q$ therefore has Euler characteristic per vertex
\begin{equation}
    \Delta\chi = 1 - \frac{q}{2} + \sum_{i=1}^q \frac{1}{n_i} = \frac{1}{2}\left(2 - \sum_{i=1}^q\frac{n_i-2}{n_i}\right).
    \label{SI:eq:Euler_characteristic_per_vertex}
\end{equation}
As an example, let us compare the Euclidean $\{3,6\}$ to the hyperbolic $\{3,7\}$ tessellation.
For the former, Supplementary Equation~(\ref{SI:eq:Euler_characteristic_per_vertex}) gives $\Delta\chi=0$, consistent with flat space, and for the latter, $\Delta\chi=-1/6<0$, consistent with hyperbolic space.

The Euler characteristic per site allows us, via the Gauss-Bonnet theorem, to compute the area per vertex $\alpha$.
The Gauss-Bonnet theorem relates the Euler characteristic to the curvature
\begin{equation}
    \int_\alpha K\dd{A} = 2\pi\Delta\chi.
\end{equation}
In the hyperbolic plane we consider, we have constant curvature $K=-4$, such that the left-hand side evaluates to $-4\alpha$ and we find
\begin{equation}
    \alpha = -\frac{\pi}{4}\left(2-\sum_{i=1}^q\frac{n_i-2}{n_i}\right).
    \label{SI:eq:area_per_vertex}
\end{equation}

In this section we study three hyperbolic tessellations of the hyperbolic plane: (i) $\{3,7\}$, (ii) $\{7,3\}$, and (iii) $6.6.7$ (also called the hyperbolic soccerball), illustrated in panels a--c of Supplementary Figure~\ref{SI:fig:tessellations:fixed_r0}.
The area per vertex for those is (i) $\pi/12$, (ii) $\pi/28$, and (iii) $\pi/84$.
We compare these three tessellations with respect to three properties:
First, in Supplementary Note~\ref{sec:approx_continuum} we consider the accuracy of approximating the continuum.
This is determined by the density of the tessellation, which in turn depends on the vertex configuration (or equivalently on $\alpha$).
Subsequently, in Supplementary Note~\ref{sec:curvature_finite_lattice}, we examine the total (i.e., integrated over the area) curvature that can be obtained in a~finite lattice with a~fixed number of vertices.
Finally, in Supplementary Note~\ref{sec:rotation_symmetry}, we discuss the effect of rotation symmetry with respect to a~central vertex on the spectrum and the profile of the eigenmodes.

\subsection{Approximating the continuum}\label{sec:approx_continuum}

\begin{figure*}
	\centering
	\includegraphics{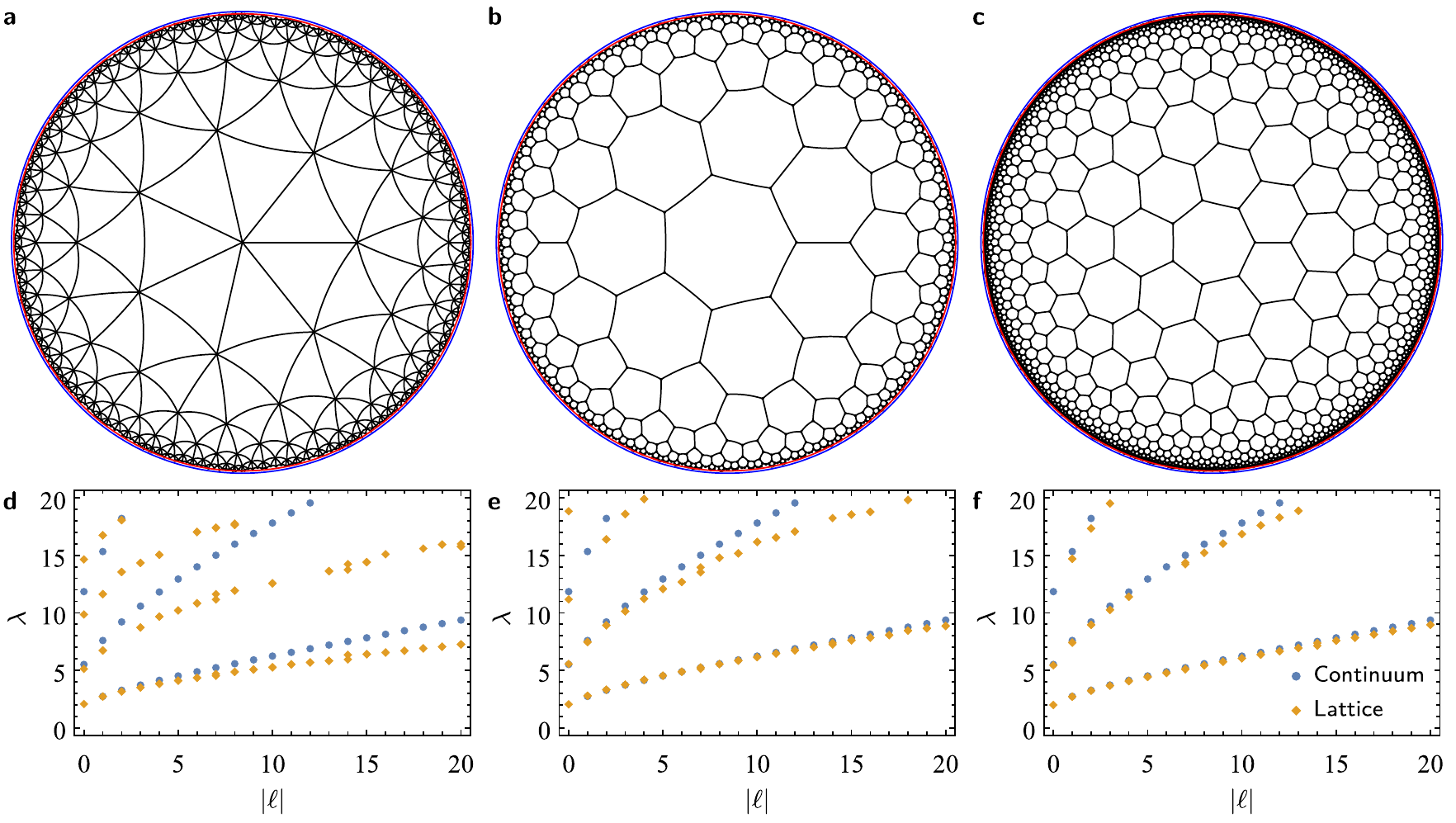}
	\titledcaption{Comparison of hyperbolic tessellations for fixed disk radius.}{
	    \textbf{a,b,c} Tessellations $\{3,7\}$, $\{7,3\}$ and $6.6.7$ of the hyperbolic plane in Poincaré disk representation (the unit circle is shown in blue), respectively, covering a~disk of radius $r_0=0.99$ (red circle). This results in $589$, $1197$, and $3857$ vertices, respectively.
	    \textbf{d,e,f} Angular momentum dispersion of the eigenstates of the graph Laplacian for the three tessellations (orange diamonds) compared to the same for the eigenmodes of the continuum Laplace-Beltrami operator on the hyperbolic drum with the same radius $r_0$ (blue disks).
	}
	\label{SI:fig:tessellations:fixed_r0}
\end{figure*}

Here, we analyze how well the three tessellations shown in Supplementary Figure~\ref{SI:fig:tessellations:fixed_r0} approximate the continuum by comparing the spectra of the graph Laplacian on the lattice to the ones of the Laplace-Beltrami operator on a corresponding disk.
All three tessellations cover approximately the same disk of radius $r_0<1$; however, due to them having different area per vertex $\alpha$, cf.~Supplementary Equation~(\ref{SI:eq:area_per_vertex}), the number of vertices varies between the three cases.
More specifically, for each of the three tessellations we compare the spectrum of the graph Laplacian $Q$ to the spectrum of the Laplace-Beltrami operator with Dirichlet boundary conditions for a~disk of the same radius $r_0$.

We have already discussed this problem analytically in Supplementary Note~\ref{sec:lattice_regularization} and have found that the graph Laplacian $Q$ is approximated by the Laplace-Beltrami operator up to corrections of order $h^3$, where $h=0.496\,970$ for $\{3,7\}$, $h=0.275\,798$ for $\{7,3\}$, and $h=0.165\,657$ for $6.6.7$.
We therefore anticipate these correction to be smallest for the $6.6.7$ tessellation, in agreement with the area per vertex being smallest for this tessellation.
We now verify this explicitly for the three tessellations on disks of radius $r_0=0.99$ by numerically computing the spectrum (eigenvalue as a~function of angular momentum) and comparing the dispersion to the one obtained from the continuum, as we did in \cref{fig:dispersion} in the main text.
Note that because of finite-size effects, our method fails to correctly identify the angular momentum of certain highly excited states (see the corresponding discussion in the Methods section).
The results are shown in Supplementary Figure~\ref{SI:fig:tessellations:fixed_r0} and we observe that the difference between lattice (orange diamonds) and continuum (blue disks) dispersion is smaller for tessellations with small area per site, i.e., when the total number of sites is larger.

\subsection{Signatures of negative curvature}\label{sec:curvature_finite_lattice}

\begin{figure*}
	\centering
	\includegraphics{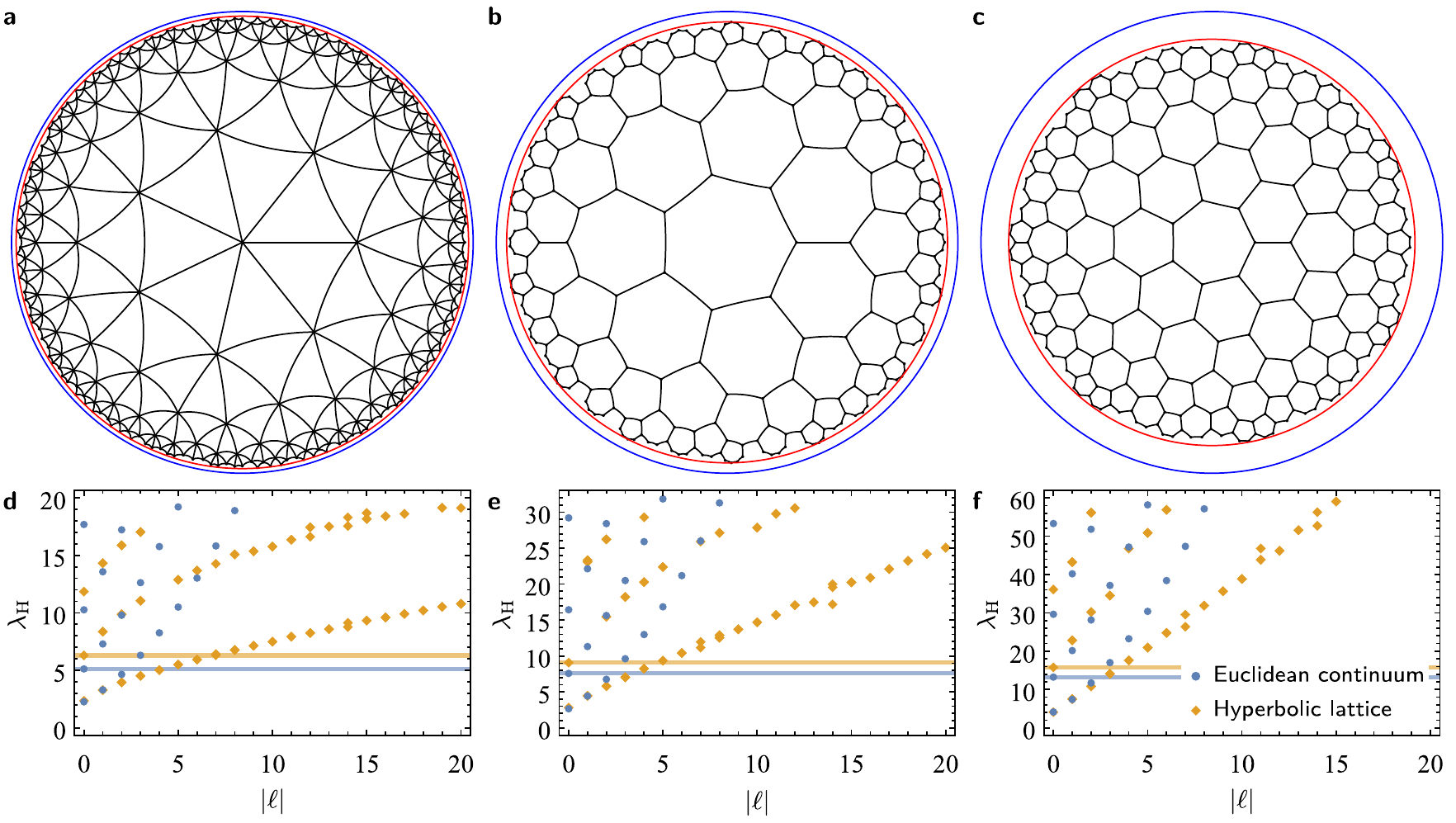}
	\titledcaption{Comparison of hyperbolic tessellations for fixed number of vertices.}{
	    \textbf{a,b,c} Tessellations $\{3,7\}$, $\{7,3\}$ and $6.6.7$ of the hyperbolic plane in Poincaré disk representation (the unit circle is shown in blue), respectively, with approximately $275$ vertices (the exact numbers of vertices are $274$, $273$ and $280$, respectively).
	    The red circle indicates the bounding circle of each tessellation with radii $r_0=0.98$, $0.955$ and $0.88$, respectively.
	    \textbf{d,e,f} Angular momentum dispersion of the eigenstates of the graph Laplacian for the three tessellations (orange diamonds) compared to the same for the eigenmodes of the continuum Laplace operator on the Euclidean drum with the same radius $r_0$ (blue disks). Horizontal lines of the corresponding color indicate the eigenvalue of the $(n,\ell)=(1,0)$ mode in each geometry. The difference in the number of modes below the orange and the blue line quantifies the spectral reordering between the hyperbolic and the Euclidean disk. Note that the eigenvalues $\lambda_\mathrm{E}$ of the Euclidean drum are rescaled and shifted to allow for a~better qualitative comparison to the hyperbolic dispersion, i.e., to emphasize the reordering of eigenstates.
	}
	\label{SI:fig:tessellations:fixed_N}
\end{figure*}

Above we have answered the question which of the three tessellations gives the best approximation of the continuum for a~disk with fixed radius $r_0<1$.
Experimentally, however, we are interested in a~different question:
For a~given number of vertices (sites), which tessellation gives the strongest signatures of negative curvature?
Naturally, we expect tessellations that cover a~larger area of the Poincaré disk to exhibit stronger signatures of the negative curvature.
Therefore, a~large area per vertex is desirable.
According to Supplementary Equation~(\ref{SI:eq:area_per_vertex}) and the values given in the paragraph following that equation, the $\{3,7\}$ tessellation is the one with the largest area per vertex out of the three under consideration.

We fix the (approximate) number of sites to 275 and construct the tessellations such that they consist of full shells.
The resulting lattices are shown in Supplementary Figure~\ref{SI:fig:tessellations:fixed_N}.
Here, we compare the angular momentum dispersion to the dispersion obtained from the eigenmodes of the continuum Laplace-Beltrami operator on the Euclidean drum of the same radius $r_0$, each.
The signature of negative curvature which we have identified in the main text, i.e., the reordering of the eigenstates compared to the Euclidean case, is with a difference of four states strongest for the $\{3,7\}$ tessellation (panels \textbf{a, d}) and reduced to only a~single state for $6.6.7$ (panels \textbf{c, f}).
Therefore, to reveal the spectral reordering in an experimental realization with a~limited number of sites, it may be desirable to opt for the $\{3,7\}$ tessellation.

\subsection{Role of rotation symmetry}\label{sec:rotation_symmetry}

Finally, we discuss the role of rotation symmetry.
The tessellations shown in Supplementary Figure~\ref{SI:fig:tessellations:fixed_r0}b,c can be shifted such that they have a~vertex at the centre of the disk.
As we argued in the main text, this is advantageous in order to to excite and detect $\ell=0$ modes which have a~maximum amplitude at the centre of the disk.
However, in the case of the $\{7,3\}$ tessellation, the seven-fold rotation symmetry is broken down to a~three-fold rotation symmetry, while in the case of the $6.6.7$ tessellation no rotation symmetry is remaining at all.
These shifted tessellations are displayed in Supplementary Figure~\ref{SI:fig:tessellations:rotation_symmetry}b,c.

In Supplementary Figure~\ref{SI:fig:tessellations:rotation_symmetry}d--f we show for each of the first 20 eigenmodes of the graph Laplacian of each considered tessellation their angular momentum $\ell$ and their weight at the central vertex.
From the continuum we expect that only $\ell=0$ modes have non-vanishing weight at that vertex.
This, indeed, holds on the $\{3,7\}$ lattice for all $\abs{\ell}\leq 6$.
After that, we observe that eigenvalues of the two $\abs{\ell}=7$ modes (and similarly for integer multiples of $7$) are split (cf.~Supplementary Figure~\ref{SI:fig:tessellations:fixed_r0}d), in stark contrast with the continuum case where such modes are degenerate.
We also observe that one of these two modes acquires a~non-zero weight at the central vertex.
An analogous feature is observed for the $\{7,3\}$ tessellation, where the modes with $\abs{\ell}$ being integer multiples of $3$ are similarly contaminated.
Finally, the situation for the $6.6.7$ tessellation is even less ideal as here most of the modes acquire a~non-vanishing weight at the central vertex.

Therefore, we conclude that a~small order of rotation symmetry leads to a~larger number of $\ell\neq 0$ eigenmodes with non-vanishing weight at the central vertex.
This in turn prevents us from easily detecting (and exciting) $\ell=0$ modes via the central vertex, as stated in the main text.

\begin{figure*}
	\centering
	\includegraphics{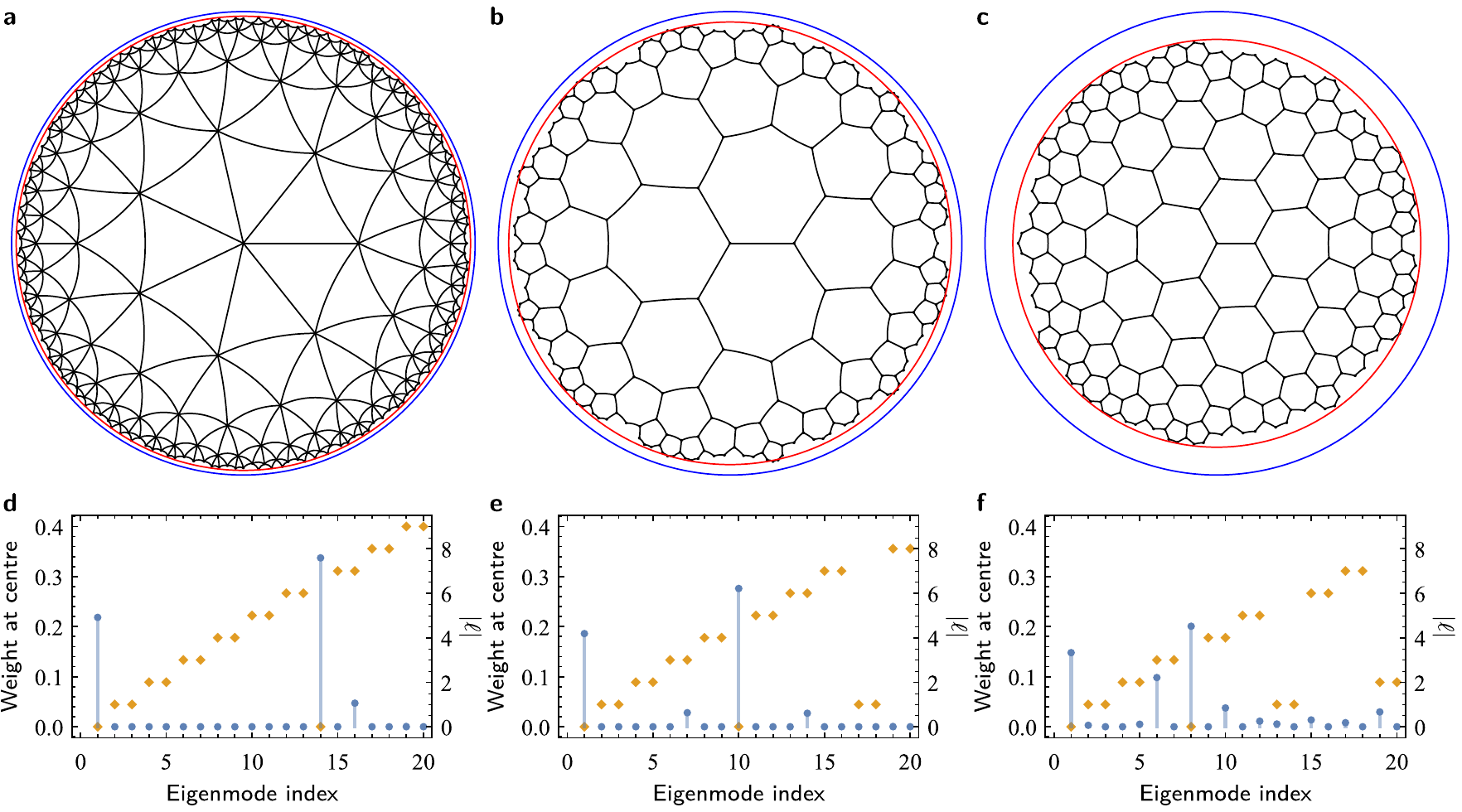}
	\titledcaption{Comparison of hyperbolic tessellations with different order of rotation symmetry.}{
	    \textbf{a,b,c} Tessellations $\{3,7\}$, $\{7,3\}$ and $6.6.7$ of the hyperbolic plane in Poincaré disk representation (the unit circle is shown in blue), respectively, with approximately $275$ vertices (the exact numbers of vertices are $274$, $271$ and $271$, respectively) and a~vertex at the centre.
	    The red circle indicates the bounding circle of each tessellation with radii $r_0=0.98$, $0.96$ and $0.88$, respectively.
	    \textbf{d,e,f} For the first 20 eigenmodes (counting degenerate modes), the absolute value of their weight at the central vertex (blue disks, left vertical axis) and absolute value of their angular momentum $\abs{\ell}$ (orange diamonds, right vertical axis) are shown.
	}
	\label{SI:fig:tessellations:rotation_symmetry}
\end{figure*}

\section{Parasitic resistances}
\begin{figure}[h]
	\hspace*{-0.75cm}
	\includegraphics[width=12cm]{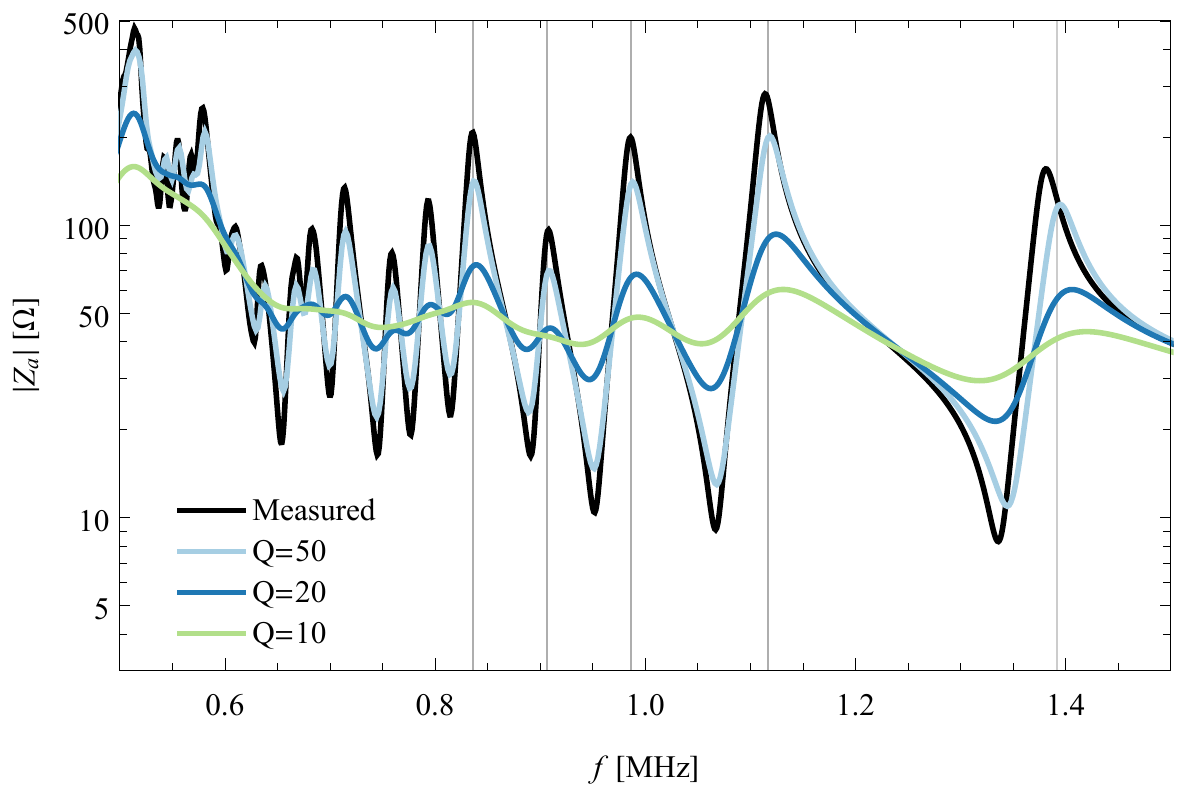}
	\caption{Comparison of the simulated impedance spectrum at node 18 for different Q factors of the inductors and measurement values.}
	\label{fig:Qfactor}
\end{figure}

Resolving individual peaks in the resonance spectrum requires a sufficiently high Q factor for all circuit elements. 
With increasing parasitic resistances, the resonance peaks of individual modes widen and flatten, making them harder to identify in an impedance sweep.
In practice, inductors are the main source of parasitic resistances in our circuit.
The Q factor of an inductor is defined through its impedance as $Q(\omega) = \frac{|Z_L(\omega)|}{\Re{Z_L(\omega)}}$.
For $Q\gg1$, we approximate $|Z_L| \approx |\mathrm{i}\omega L|$, and obtain $Z_L=\omega(\mathrm{i}+1/Q)L$.
Supplementary Figure~\ref{fig:Qfactor} compares simulated impedance sweeps of node 18 for several constant Q factors of the inductors and the measured values.
For Q factors of 50 and 20, all relevant impedance peaks can be easily identified, while at a Q factor of 10, the Peak at $0.906$ MHz is no longer recognizable.

Supplementary Figure~\ref{fig:Qfactor} shows that the measured data is consistent with $ Q>50 $ in the measured frequency interval. For a given eigenvector $V_n$ of the hopping matrix $M$ with corresponding eigenvalue $\lambda_n$, the Laplacian equation
\begin{equation}
    \left(\mathrm{i} \,\omega \,C \,M + \frac{1}{\omega(\i+1/Q)L} \mathds{1}\right) V_n = 0   
\end{equation}
reduces to the scalar equation
\begin{align}
\mathrm{i}\omega C \lambda_n + \frac{1}{\omega(\mathrm{i}+1/Q)L} = 0,   
\end{align}
which is equivalent to that of a simple serial R-L-C oscillator. Note that in this equation, $M$ denotes the hopping matrix, since $Q$ is already used for the $Q$-factor. Since real and imaginary part of a serial oscillator's eigenfrequency are related by $\Re{\omega_0}/\Im{\omega_0}\approx 2Q$, we expect decay times of free oscillations in the circuit network to exceed 100 oscillation periods.

\section{Extended Analysis of Measured Eigenmodes}
In this section we discuss additional data on the measured eigenmodes and perform an extended comparison to theory.
We present extended versions of the right panel of \cref{fig:dispersion}b and \cref{fig:experiment}c in the main text, and we quantitatively analyze the deviations of the experimentally extracted data from the theoretical prediction based on the Laplacian matrix of the hyperbolic lattice.

\begin{figure*}
	\centering
	\includegraphics{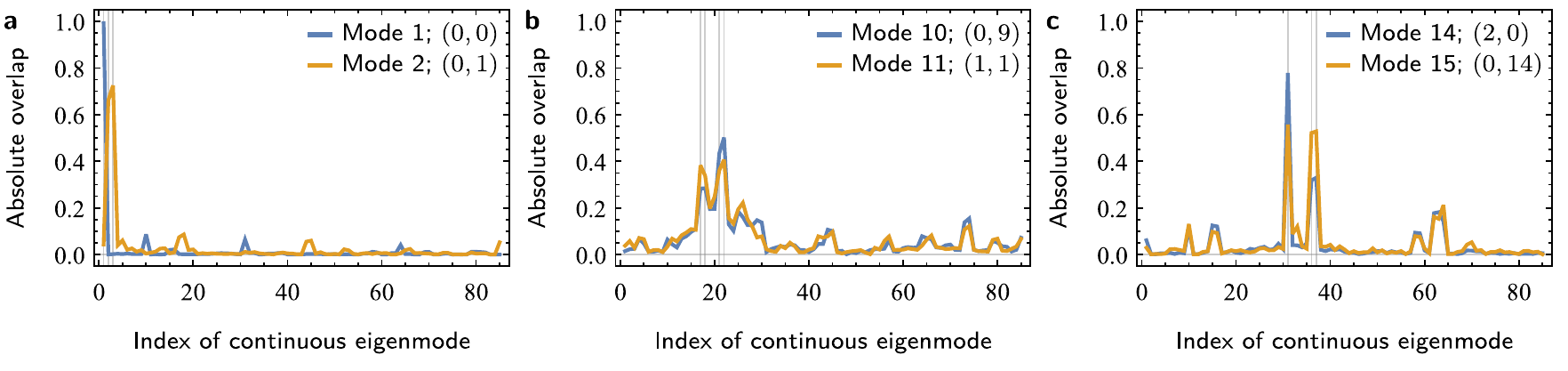}
	\titledcaption{Identifying the quantum numbers of the measured eigenmodes.}{
	    Absolute value of the overlap of some of the measured modes with the eigenmodes of the Laplace-Beltrami operator with Dirichlet boundary conditions on a disk of radius $r_0=0.94$.
	    The vertical gray lines indicate the positions of the maxima of the overlap, which allow us to assign quantum numbers $(n,\abs{\ell})$ to each mode (see legend).
	    \textbf{a} The low-energy modes (corresponding to small $\lambda$) can be easily matched to continuous eigenmodes because they only show significant overlap with a single mode with $\ell=0$ (e.g., mode 1; blue line) or a pair of modes with $\pm\ell$ (e.g., mode 2; orange line).
	    For some specific modes, this is not the case:
	    \textbf{b} Modes 10 and 11 both have significant overlap with $(n,\abs{\ell})=(0,9)$ and $(1,1)$, which can be understood from the fact that these two modes are close to being accidentally degenerate (Supplementary Figure~\ref{SI:fig:eigenmodes:quantitative_comparison}a).
	    \textbf{c} The breaking of the continuous rotation symmetry by the lattice lifts the degeneracy of the $(1,-14)$ and $(1,+14)$ mode such that one of them ends up close in eigenvalue to the $(2,0)$ mode, leading to their hybridization and to a significant overlap with the corresponding continuous eigenmodes.
	}
	\label{SI:fig:eigenmodes:LBdecomp}
\end{figure*}

To compare the experimental results to theoretical predictions, the first step is to match the eigenmodes.
In the main text, we have used a simple Fourier transform on the outermost sites to determine the angular momentum $\ell$ and match the modes according to the quantum numbers $n$, $\ell$.
As discussed at that point, such a procedure works well for the lowest couple of eigenmodes, but becomes increasingly inaccurate with increasing $n$ and $\ell$.
To analyze more modes, we therefore use the alternative method described in Supplementary Note~\ref{sec:approx_continuum}.
Recall that there we theoretically analyzed the overlap of the graph Laplacian's eigenvectors to the discretized eigenmodes of the continuum Laplace-Beltrami operator (on the appropriate disk and with appropriate boundary conditions).
We repeat here an analogous analysis for the experimentally extracted mode profiles (see Supplementary Figure~\ref{SI:fig:eigenmodes:LBdecomp} for several examples) and determine $(n,\abs{\ell})$ by identifying the continuous eigenmode for which the overlap is largest (see Supplementary Figure~\ref{SI:fig:eigenmodes:mode_comparison} for the results).

\begin{figure*}
	\centering
	\includegraphics{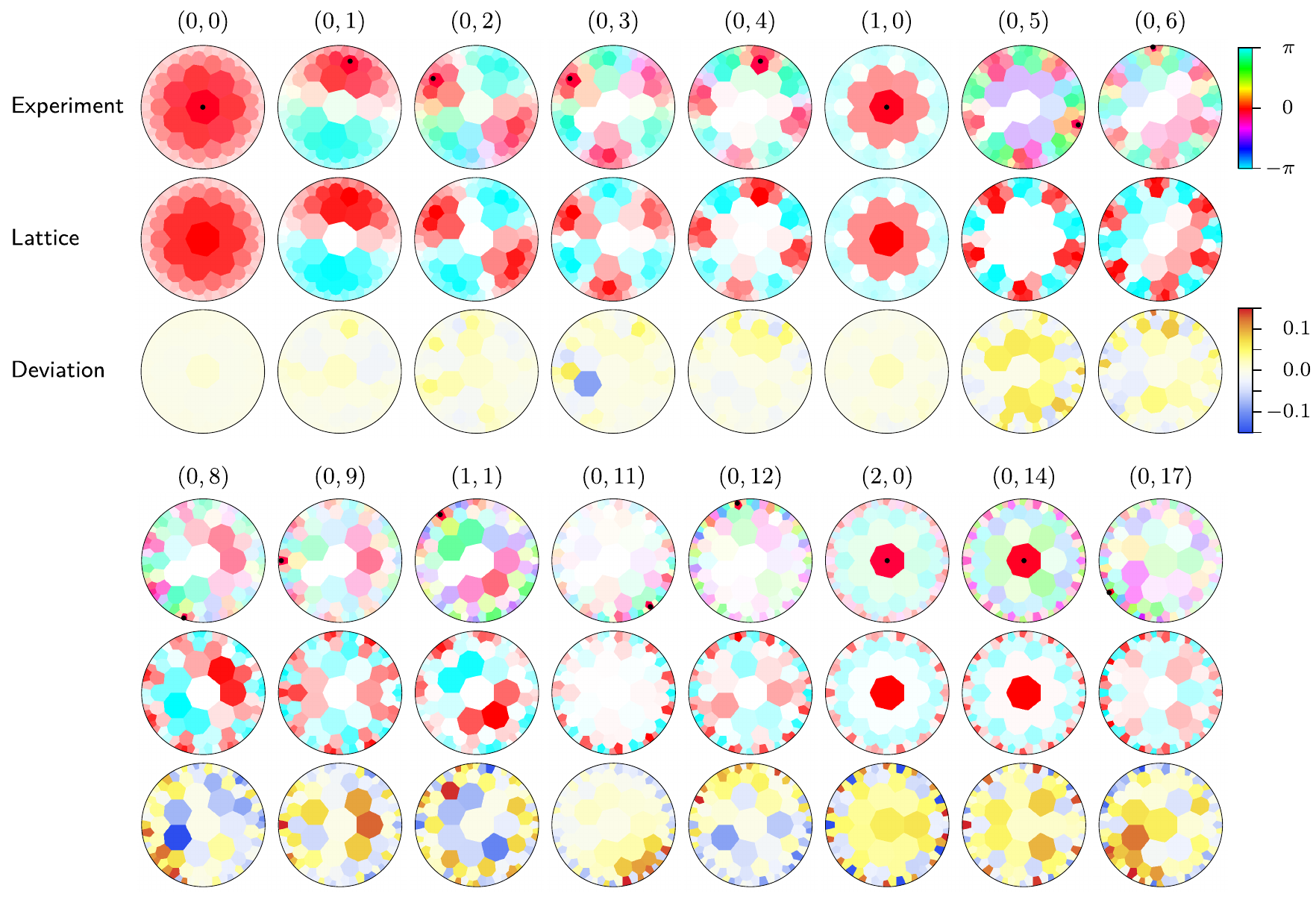}
	\titledcaption{Comparison of measured eigenmodes to eigenvectors of the graph Laplacian matrix.}{
	    The three rows (continued in the lower half of the figure) show the following data.
	    \textbf{Experiment}: the voltage profile of the measured eigenmodes with saturation encoding the magnitude as a fraction of the voltage (white denotes $0$ and full saturation $1$) at the input node (black dots) and color encoding the phase relative to the reference voltage (see legend on the right).
	    \textbf{Lattice}: eigenvectors obtained from diagonalizing the Laplacian matrix defined by hyperbolic lattice (saturation and color as for the experiment).
	    \textbf{Deviation}: difference between the normalized experimental data and the data on the lattice (see legend on the right).
	    For each mode the quantum numbers $n$ and $\abs{\ell}$ are extracted by determining the continuous eigenmode with maximal overlap (see Supplementary Figure~\ref{SI:fig:eigenmodes:LBdecomp}) and given at the top in the format $(n,\abs{\ell})$. The modes from experiment and theory are matched according to those numbers.
	}
	\label{SI:fig:eigenmodes:mode_comparison}
\end{figure*}

In Supplementary Figure~\ref{SI:fig:eigenmodes:mode_comparison} we further compare the $16$ eigenmodes that we were able to excite, measure, and identify successfully with the corresponding eigenvectors of the adjacency matrix obtained numerically from the theoretical hyperbolic lattice.
The deviation from theory is quantified by (1)~the point-wise difference and by (2)~the overlap.
The former is plotted in Supplementary Figure~\ref{SI:fig:eigenmodes:mode_comparison} and analyzed in Supplementary Figure~\ref{SI:fig:eigenmodes:quantitative_comparison}b, which shows the mean and standard deviation of the point-wise difference between experiment and theory (green squares).
Supplementary Figure~\ref{SI:fig:eigenmodes:quantitative_comparison}b also shows the overlap of the experimentally measured eigenmodes with the theoretical predictions (red triangles).
Furthermore, we compare the measured and predicted eigenvalues in Supplementary Figure~\ref{SI:fig:eigenmodes:quantitative_comparison}a (an extended version of the right panel of \cref{fig:dispersion}b in the main text), i.e., the eigenvalue as a function of $\abs{\ell}$ for both the experimental as well as theoretical (lattice) data with relative errors shown in the inset.

\begin{figure*}
	\centering
	\includegraphics{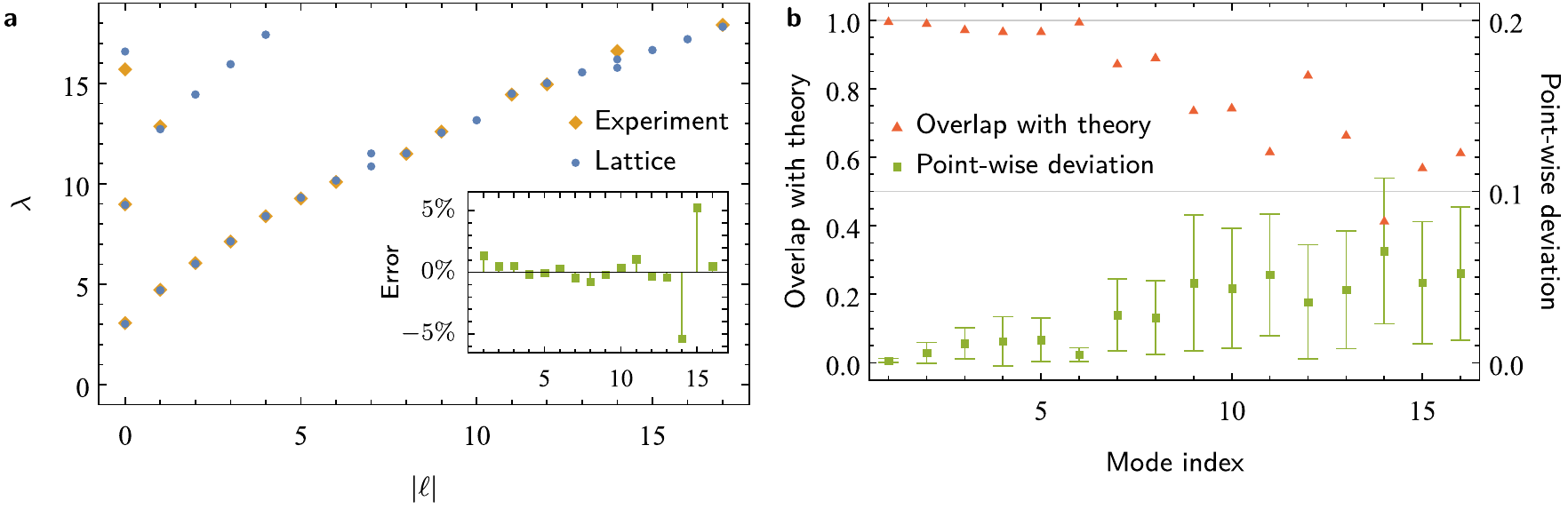}
	\titledcaption{Quantitative comparison measured eigenmodes and eigenvalues to theory.}{
	    \textbf{a} Angular momentum dispersion, i.e., eigenvalue $\lambda$ vs.\ the absolute value of the angular momentum $\ell$ for each eigenmode. Data obtained from measurements of the electric circuit (orange diamonds) and from diagonalizing the Laplacian matrix defined by the hyperbolic lattice (blue disks) are shown. The inset shows the relative error in the experimental data compared to the theoretical prediction (horizontal axis: mode index according to increasing $\lambda$ in the experimental data); there are only two outliers (modes $14$ and $15$) with a relative error significantly larger than $1\%$ (see text).
	    \textbf{b} Comparison of the experimentally extracted eigenmodes to the eigenvector of the Laplacian matrix (both shown in Supplementary Figure~\ref{SI:fig:eigenmodes:mode_comparison}).
	    The red triangles show the absolute overlap of the experimentally and theoretically obtained eigenvectors and the green squares the mean of the absolute value of the point-wise deviation (interval marks indicate the standard deviation computed over all the nodes in the circuit).
	    Again, mode $14$ can be identified as an outlier (see text).
	}
	\label{SI:fig:eigenmodes:quantitative_comparison}
\end{figure*}

We observe that the relative error for almost all of the $16$ eigenmodes is below $2\%$ (cf.~Supplementary Figure~\ref{SI:fig:eigenmodes:quantitative_comparison}a); the two outliers, modes $14$ and $15$, are discussed below.
In contrast, Supplementary Figure~\ref{SI:fig:eigenmodes:quantitative_comparison}b shows that the measured eigenmodes agree very well with theory for modes $1$ to $6$ after which the deviations start to increase.
This is reflected both in the overlap of the measured modes with the theoretically expected ones, as well as in the point-wise deviations at the individual circuit nodes (plotted also in Supplementary Figure~\ref{SI:fig:eigenmodes:mode_comparison}).
Note that additionally, due to parasitic effects, the experimental data shows increasing deviations in the phase compared to the theory where only $0,\pi$ phases occur (cf.~Supplementary Figure~\ref{SI:fig:eigenmodes:mode_comparison}).

The deviations in $\lambda$, i.e., in the eigenfrequencies of the circuit, are weakly dependent on the index of the excited modes (except for the outliers mentioned above).
This indicates that these deviations most likely can be attributed to parasitic effects and to disorder in the circuit components, as these are both expected to exhibit such a weak dependence on the index of the excited modes.
The eigenfrequencies are extracted from impedance measurements such as \cref{fig:experiment}b in the main text; as long as the peaks are well separated, they can be accurately measured.
Note that problems can arise when two modes are close to each other in eigenvalue, i.e., almost accidentally degenerate, which is the case for modes 10 and 11.
Supplementary Figure~\ref{SI:fig:eigenmodes:LBdecomp}b shows that both modes have significant overlap with the $(n,\abs{\ell})=(0,9)$ and $(1,1)$ eigenmodes of the continuum Laplace-Beltrami operator.
This makes it more challenging to excite and assign quantum numbers to those modes.
On the other hand, the mode profiles are much more sensitive to other error sources.
First, in the experiment, it is impossible to excite exactly a single mode, generally a superposition of several modes is excited.
When the eigenmodes are well separated in frequency or if the input node lies in a nodal plane of many other eigenmodes, the additional eigenmodes have a small weight in the superposition.
However, with increasing mode number the frequency separation is reduced, such that the deviations from theory increase gradually.

Finally, we comment on the missing data for the $\abs{\ell}=7$ mode as well as the large difference of the experimentally extracted and theoretically predicted eigenvalues for the $(n,\abs{\ell})=(2,0)$ and $(1,14)$ modes (modes $14$ and $15$ have a relative error that is significantly larger than the typical error, cf.~Supplementary Figure~\ref{SI:fig:eigenmodes:quantitative_comparison}a).
As discussed in Supplementary Note~\ref{sec:rotation_symmetry}, for rotation symmetry of finite order some modes with $\ell\neq 0$ attain a non-vanishing amplitude at the origin and the degeneracy with the second mode with identical $n,\abs{\ell}$ is lifted (only one of the two modes has a significant amplitude at the origin).
We have not managed to cleanly excite the $(1,7)$ modes, but the described phenomenon is visible in the data for the $(1,14)$ mode (cf.~Supplementary Figure~\ref{SI:fig:eigenmodes:mode_comparison}).
This also allows us to understand the large deviation of the eigenvalues of the $(2,0)$ and $(1,14)$ modes: both have significant weight at the origin (cf.~Supplementary Figure~\ref{SI:fig:eigenmodes:mode_comparison}) and eigenvalues that are expected to be very close to each other (cf.~Supplementary Figure~\ref{SI:fig:eigenmodes:quantitative_comparison}a).
Therefore, it is difficult to excite only one of them, such that the experimentally excited modes are very likely superpositions of the two.
This is reflected in the overlap of the measured modes with the eigenmodes of the continuum Laplace-Beltrami operator in Supplementary Figure~\ref{SI:fig:eigenmodes:LBdecomp}c.
While based on the voltage profile (either visually or via the Fourier transform) the lower mode is identified as the $\ell=0$ mode, the eigenvalues would suggest the opposite (cf.~Supplementary Figure~\ref{SI:fig:eigenmodes:quantitative_comparison}a).
Choosing different input nodes should resolve this issue.

In conclusion, we find that getting accurate data on the eigenvalues is not a problem as long as the modes can be cleanly excited, while the error in the eigenmode profiles increases gradually with increasing mode number.
Nevertheless, correctly identifying the modes remains possible even with reduced accuracy of the mode profiles.
To recognize the reordering of the eigenmode compared to flat space, the accuracy of our experimental setup is more than sufficient: only the first six modes are required, for which the overlap with theory is above $95\%$; but we have demonstrated that, already without any additional optimization, it is possible to find good agreement with theory for higher modes as well.

\section{Signal propagation in the electric circuit network}\label{sec:signal_propagation}

In this section we briefly explain the time-dependent behavior of the hyperbolic circles of constant phases in \phasefront{}.
While we have already discussed the results at fixed times in the main text, the time-dependence requires some additional explanation.
In particular, we observe that the hyperbolic circles of constant phases are falling into the input node.
This is a~consequence of our specific electric circuit network being a~negative-index metamaterial (also called left-handed), i.e., having a~negative refractive index.

The propagation of waves on a~drum is generally given by the following differential equation involving the Laplace-Beltrami operator:
\begin{equation}
    \frac{1}{c^2}\pdv[2]{}{t}u(t,x,y)-\Delta_\mathrm{g}u(t,x,y) = 0
\end{equation}
with the wave speed $c$ determined by the medium.
In the Euclidean case, for example, the wave-equation leads to the dispersion $\omega(k) = c\abs{\vec{k}}$ with the two-dimensional momentum vector $\vec{k}$; thus, phase and group velocity are equal: $v_p=v_g=c$.
The situation in the experiment corresponds to an additional inhomogenous source term $S(x,y,t)\sin(\omega t)$, where $S$ is localized both in space (where the excitation happens) and in time (pulse-like), i.e.,
\begin{equation}
    \frac{1}{c^2}\pdv[2]{}{t}u(t,x,y)-\Delta_\mathrm{g}u(t,x,y) = S(x,y,t)\sin(\omega t).
\end{equation}
The source term leads to an excitation of eigenmodes of the drum, i.e., of $-\Delta_\mathrm{g}$, according to its frequency spectrum and the pulse propagates across the drum with speed $c$.

For an electric circuit there are some important differences to the dynamics, even though the situation is conceptually the same.
According to Kirchhoff's law, the differential equation governing our electric circuit network is
\begin{equation}
    \pdv{}{t}I_a = CQ_{ab}\pdv[2]{}{t}V_b-\frac{1}{L}V_a,
\end{equation}
where $I_a(t)$ and $V_a(t)$ are the input current and voltage at node $a$, $C$ the capacitance coupling two adjacent nodes, $L$ the inductance to ground for each node and $Q_{ab}$ the graph Laplacian describing the (capacitive) connections between the nodes.
The continuum limit therefore is
\begin{equation}
    C\pdv[2]{}{t}\Delta_\mathrm{g}V(t,x,y) - \frac{1}{L}V(t,x,y) = \pdv{}{t}I(t,x,y)
\end{equation}
where the voltage field $V(t,x,y)$ takes the role of $u(t,x,y)$ and the time-derivative of the input current $\pdv{}{t}I(t,x,y)$ the role of the source.

The modified wave-equation
\begin{equation}
    LC\pdv[2]{}{t}\Delta_\mathrm{g}u(t,x,y) - u(t,x,y) = 0
\end{equation}
results in the group velocity $v_g$ having the opposite sign compared to the phase velocity $v_p$.
In the Euclidean case, for example, the dispersion is
\begin{equation}
    \omega(k) = \frac{1}{\sqrt{LC}}\frac{1}{k},
\end{equation}
which implies
\begin{equation}
    v_g = \dv{\omega}{k} = -\frac{1}{\sqrt{LC}}\frac{1}{k^2} = -v_p.
\end{equation}
This explains the observation in \phasefront{} that the hyperbolic circles of constant phase seem to fall into the input node (with $v_p$), while the excited voltage pulse is propagating away from it (with $v_g$).

\let\oldaddcontentsline\addcontentsline     
\renewcommand{\addcontentsline}[3]{}        

\titleformat{\section}
  {\centering\normalfont\fontsize{10pt}{11pt}\selectfont\bfseries\selectfont\uppercase}{\uppercase{Supplementary Movie~\thesection.}}{1em}{}

\setcounter{section}{0}

\pdfbookmark[0]{Supplementary Movie 1}{supp-movie}
\section{Measured signal propagation in the electric circuit}
Application of a short and spatially localized pulse applied to node 31 (blue curve in the left panel) leads to a wave propagating through the circuit. The voltage response at node 31 is shown as an orange curve in the left panel and the instantaneous phase at each node in the right panel. The nodes are indicated by black dots, and concentric hyperbolic circles with center at node 31 are shown in black to illustrate the hyperbolic metric.

\pdfbookmark[0]{References}{supp-references}
\putbib

\end{bibunit}

\renewcommand{\bibliography}{}

\end{document}